\documentclass[12pt, draftclsnofoot, onecolumn]{IEEEtran}
\usepackage{tikz}
\usepackage{float}
\usepackage{amsmath}
\usepackage{amssymb}
\usepackage{commath}
\usepackage{algorithm}
\usepackage{algorithmic}
\usepackage{multicol}
\usepackage[noadjust]{cite}
\usepackage{balance}
\usepackage{pgfplots}
\usepackage{ragged2e}
\usetikzlibrary{arrows,arrows.meta,
decorations.markings,snakes,patterns,positioning,fit,shapes}
\newcommand*{\argmax}{\ensuremath{\mathop{\mathrm{arg\,max}}}}
\newcommand{\dij}[2]{\sqrt{(x_i^{#2} - x_{#1})^2 + (y_i^{#2} - y_{#1})^2}}

\newcommand{\diag}{\text{diag}}
\newcommand{\tijt}[1]{\tilde{T}_{ij}^{#1}}
\newcommand{\rijt}[1]{\tilde{R}_{ij}^{#1}}
\newcommand{\tij}[1]{T_{ij}^{#1}}
\newcommand{\rij}[1]{R_{ij}^{#1}}

\newcommand{\ttat}[1]{\boldsymbol{\xi}_i^{#1}}
\newcommand{\tgam}{\frac{1}{\tilde{\gamma}_i}}
\newcommand{\tg}{\tilde{\gamma}_i}
\newcommand{\ttet}{\tilde{\theta}_i}
\newcommand{\pos}{\mathbf{p}_i}

\newcommand{\tetvec}[1]{\boldsymbol{\tilde{\vartheta}}_{#1}}
\newcommand{\tetvect}[1]{\boldsymbol{\xi}_{#1}}

\newcommand{\zji}{\mathbf{z}_{ij}}

\newcommand{\ccij}{\mathbf{c}_{ij}}

\newcommand{\cntnt}[2]{p(\ccij^{#1}, \varphi_{ij}^{#1},  \zeta_{ij}^{#1}|\tetvect{i}^{#2})}
\newcommand{\tnt}[2]{p(\tetvect{i}^{#1}|\tetvect{i}^{#2})}
\newcommand{\tntt}[2]{p(\tetvect{i}^{#1}|\tetvect{i}^{#2})}

\newcommand{\tta}[1]{\boldsymbol{\xi}_i^{#1}}

\newcommand{\cn}{\hat{p}_{\text{nlos}}}

\newcommand{\mug}[1]{\boldsymbol{\mu}({#1})}
\newcommand{\mugg}[1]{\boldsymbol{\mu}({#1})_f^k}
\newcommand{\sigg}[1]{\boldsymbol{\Sigma}({#1})}

\newcommand{\qn}[1]{\mathbf{Q}_n({#1})}
\newcommand{\nant}{N_{\text{ant}}}

\newcommand{\abb}[2]{\textbf{#1: }#2}
\begin{document}
\title{DNN-assisted Particle-based Bayesian Joint Synchronization and Localization}

\author{Meysam Goodarzi, Vladica Sark,
Nebojsa Maletic,~\IEEEmembership{Member,~IEEE,} Jes\'us Guti\'errez,~\IEEEmembership{Member,~IEEE,} Giuseppe Caire,~\IEEEmembership{Fellow,~IEEE,} and Eckhard Grass
\thanks{M. Goodarzi, V. Sark, N. Maletic and Jes\'us Guti\'errez, and E. Grass are with IHP -- Leibniz-Institut f\"{u}r innovative Mikroelektronik, Frankfurt (Oder), Germany (emails: \{goodarzi, sark, maletic, teran, grass\}$@$ihp-microelectronics.com). M.~Goodarzi and E. Grass are also with Humboldt University of Berlin, Berlin, Germany. G. Caire is with Technical University of Berlin, Berlin, Germany (email: caire$@$tu-berlin.de).}
}

\maketitle
\begin{abstract}
In this work, we propose a Deep neural network-assisted Particle Filter-based (DePF) approach to address the Mobile User (MU) joint synchronization and localization (sync\&loc) problem in ultra-dense networks. In particular, DePF deploys an asymmetric time-stamp exchange mechanism between the MUs and the Access Points (APs), which,  traditionally, provides us with information about the MUs' clock offset and skew. However, information about the distance between an AP and an MU is also intrinsic to the propagation delay experienced by the exchanged time-stamps. In addition, to estimate the angle of arrival of the received synchronization packets, DePF draws on the multiple signal classification algorithm that is fed with the Channel Impulse Response (CIR) experienced by the sync packets. The CIR is also leveraged to determine the link condition, i.e. Line-of-Sight (LoS) or Non-LoS. Finally, to perform joint sync\&loc, DePF capitalizes on particle Gaussian mixtures that allow for a hybrid particle-based and parametric Bayesian Recursive Filtering (BRF) fusion of the aforementioned pieces of information and, thus, jointly estimates the position and clock parameters of the MUs. The simulation results verify the superiority of the proposed algorithm over the state-of-the-art schemes, especially that of the extended Kalman filter- and linearized BRF-based joint sync\&loc. In particular, only drawing on the synchronization time-stamp exchange and CIRs from a single AP, for 90$\%$ of the cases, the absolute position and clock offset estimation error remain below 1 meter and 2 nanoseconds, respectively.

\end{abstract}
%
\IEEEpeerreviewmaketitle
\section{Introduction}\label{sec:intro}
The fifth-generation (5G) of mobile communication networks is expected to deploy Access Points (APs) with a high spatial density to meet the increasing demand for mobile data traffic. As a result, Mobile Users (MUs) are expected to be most of the time in Line-of-Sight (LoS) of several APs. This also lays the ground for an accurate MU localization, which is in particular of crucial importance for services such as user tracking and location-assisted beamforming \cite{wu2010clock, maletic2018device}. Furthermore, such APs are likely to be equipped with antenna arrays and they are expected to support Fine Time Measurement (FTM) capability introduced in several standards, e.g., IEEE 802.11  \cite{7786995}. The former facilitates the Angle of Arrival (AoA) estimation, while the latter allows for the AP-MU \mbox{time-stamp} exchange, by means of which synchronization and distance measurements are enabled. The synchronization itself also plays a decisive role when performing time-based localization. In particular, for many of the \mbox{state-of-the-art} MU localization techniques to function, the clock parameters of the MUs need to be known (or to be continuously tracked). Therefore, it appears that the MU's clock parameter estimation and MU localization are closely intertwined and need to be tackled jointly.

The joint MU synchronization and localization (sync\&loc) problem has been extensively addressed in the literature \cite{zheng2009joint, yuan2016cooperative, etzlinger2017cooperative, meyer2018scalable, werner2015joint, koivisto2017joint}. The authors in \cite{yuan2016cooperative} rely on symmetric inter-agent (AP-MU, inter-MU, and inter-AP) time-stamp exchange and Belief Propagation (BP) to jointly estimate MUs' locations and clock offsets. A similar approach has been adopted by \cite{etzlinger2017cooperative,meyer2018scalable} with the aid of asymmetric \mbox{time-stamp} exchange mechanism proposed in \cite{chepuri2012joint}. While \mbox{time-stamp} exchange is expected to be supported in 5G networks \cite{7786995}, the high number of message-passings required by BP renders the approach limited in practice. Additionally, \cite{yuan2016cooperative,etzlinger2017cooperative,meyer2018scalable,chepuri2012joint,7786995} provide the estimation of the sync\&loc parameters at MUs, whereas for the location-based services to be delivered, these parameters need to be computed at the network side. Another drawback is the strong assumption of a fully cooperative network (also made in \cite{vaghefi2015cooperative}). That is, in addition to inter-AP and AP-MU communications, the MUs can also communicate with each other, which is not envisioned in 5G mobile networks. Nevertheless, the cooperation capability between the APs and the Base Stations (BSs) can be drawn on to perform hybrid synchronization as done in \cite{goodarzi2020hybrid, goodarzi2020synchronization}, laying the ground for an accurate MU joint sync\&loc.

Moreover, in \cite{werner2015joint, koivisto2017joint}, the authors leverage  Extended Kalman Filtering (EKF) to obtain the estimation of clock and position parameters in ultra-dense networks. In particular, they assume accurate inter-AP synchronization and perform MU joint sync\&loc in the presence of uncertainty about the time of arrival and AoA parameters. The level of uncertainty is then determined based on the derived Cramer-Rao bound. Another approach called Linearized BRF (L-BRF), which is based on linearizing the filter, has been employed in \cite{goodarzi2020bayesian, goodarzi2021synchronization}, albeit the perfect inter-AP synchronization assumption is lifted. Instead, the APs and their backhauling BSs are assumed to be synchronized using cooperative hybrid synchronization introduced in \cite{goodarzi2020hybrid, goodarzi2020synchronization}. While EKF and L-BRF can partially mitigate the destructive impact of nonlinearities in the measurements, they are likely to diverge if a reliable estimate of the initial state is not available \cite{stordal2011bridging}. Another weakness of these filters is the underestimation of the covariance matrix. A promising approach, on the one hand, to avoid such shortcomings of EKF/L-BRF, and, on the other hand, to boost the accuracy of position estimation, is estimating the (prediction, measurement likelihood, and posterior) distributions by means of Particle Gaussian Mixture (PGM) filters introduced in \cite{alspach1972nonlinear}. Specifically, in this approach, instead of approximating each distribution as a single Gaussian function, they are approximated with a sum weighted of Gaussian functions, or, alternatively, Gaussian mixtures \cite{gustafsson2010particle}. Nevertheless, the problem that immediately arises when using PGM filters is dimensionality, rendering the approach computationally expensive for multi-variable estimations. To overcome this drawback, we resort to a hybrid parametric and particle-based approach where we capitalize on the linear relations between the measurements and the clock parameters to reduce the dimensionality. In comparison to the standard PF, this approach features a strictly lower estimation variance as a result of Rao-Blackwell's lemma discussed in \cite{doucet2013rao}, and leads to more accurate estimates given the same number of particles \cite{gustafsson2010particle}. Specifically, PGM's performance stands out when the uncertainty increases.

Even the PGM-based localization techniques can suffer from divergence under certain conditions, e.g., improper tuning of the filter's hyper-parameters and faulty measurements, resulting mostly from Non-Line-of-Sight (NLoS) links \cite{guvenc2009survey}.
The former must be addressed when designing the filter, while the latter can be dealt with using NLoS mitigation methods such as those proposed in \cite{qi2006time, li2019joint, yin2013and}. The technique in \cite{qi2006time} relies on the multipath components of NLoS links to enhance the positioning accuracy. However, such a method functions well only in the presence of strong multipath components and prior statistics on NLoS-induced errors. The latter are also estimated and utilized along with trajectory tracking in \cite{li2019joint} to perform indoor positioning. The authors in \cite{yin2013and}, however, take another approach and model the measurement noise by a two-mode mixture distribution and approximate the maximum likelihood estimator using expectation maximization. Such approaches add an extra computation overhead that may not be necessary in dense networks where the LoS probability is around $0.8$ and increases with the AP density \cite{mondal20153d}. Therefore, to mitigate the estimation inaccuracy stemming from the faulty measurements, we draw on NLoS identification techniques to identify the NLoS links and discard them. Such an approach boosts the accuracy and features less complexity compared to the methods proposed in \cite{qi2006time, li2019joint, yin2013and}.

There is a wide spectrum of NLoS identification approaches adopted in the literature, e.g., hypothesis testing as in \cite{yu2013nlos}, the statistical approach taken in \cite{venkatraman2002statistical}, and Machine Learning (ML)-based methods such as that of \cite{marano2010nlos}. However, recently ML algorithms, particularly DNN-based approaches such as AmpN \cite{xiao2017ampn}, have drawn substantial attention in classification problems. In particular, DNNs exhibit a remarkable performance due to their ability, on one hand, in implementing almost any classifier function, and, on the other hand, in extracting task-related features from the input data \cite{heaton2008introduction, liu2017survey}. Other approaches such as Support Vector Machine (SVM), or Bayesian sequential testing require human intervention that may be, given the limited intuition, flawed, and erroneous. Furthermore, DNN units are also expected to be part of the communication devices as they are the cornerstone of many solutions for different communication problems such as slice management and anomaly detection \cite{5gc2021}. Therefore, a DNN-based NLoS identifier appears to be a reasonable choice. The input to the DNN can be signals containing class-relevant features such as received signal strength or Channel Impulse Response (CIR). The CIR turns out to be more informative about the communication environment and link condition. Therefore, for the sake of prediction accuracy, we rely on AP-MU CIRs in this work.

In addition to NLoS-identification, the CIR can also be fed into one of the \mbox{state-of-the-art} AoA estimation algorithms to obtain the signal's direction of arrival. AoA estimation has been extensively investigated in the literature. Algorithms such as MUSIC \cite{schmidt1986multiple}, reduced-dimension MUSIC \cite{zhang2010direction}, and ESPRIT \cite{roy1989esprit} can accurately estimate the AoA. A detailed comparison between them has been conducted in \cite{oumar2012comparison} concluding that the difference is negligible, albeit MUSIC slightly outperforms the others and, therefore, it is employed for the purpose of this work.

In this paper, we propose a DNN-assisted PF-based (DePF) joint sync\&loc algorithm that draws on the CIR to estimate the AoA and to determine the link condition, i.e., LoS or NLoS, thereby excluding the faulty measurements to enable a more precise parameter estimation. It then estimates the joint probability distribution of MU's clock and position parameters using a PGM filter. The dimension of the PGM filter is then reduced by revealing and exploiting the existing linear sub-structures in the measurements, thereby tackling the dimensionality problem. To the best of our knowledge, this is the first work employing a PGM filter in a hybrid particle-based and parametric manner to perform joint sync\&loc.

The contribution of this paper is summarized as follows:
\begin{itemize}
\item We present and discuss the principles of asymmetric \mbox{time-stamp} exchange and AoA estimation. The former assists in the estimation of the clock skew, offset, and the AP-MU distance, while the latter aids in the position estimation by providing the direction of an MU relative to the position of its serving APs.
\item We develop a DNN for NLoS identification based on AP-MU CIRs. The outcome of such a DNN helps to identify erroneous measurements, i.e., time-stamps and AoAs, and discard them, thereby preventing large errors in the estimation.
\item We propose a DNN-assisted particle filter-based joint sync\&loc algorithm that estimates the clock parameters and the position of an MU in a hybrid parametric and particle-based manner. Such an approach not only boosts the estimation accuracy but also overcomes the dimensionality problem that arises in particle Gaussian mixture filters due to the high number of parameters.
\item We analyze the performance of the proposed approach with the aid of detailed simulations in a challenging \mbox{real-world} scenario. In particular, the MUs' movement profile comprises acceleration, deceleration, and constant speed. Furthermore, the APs are distributed to provide signal coverage for the MUs.
\end{itemize} 
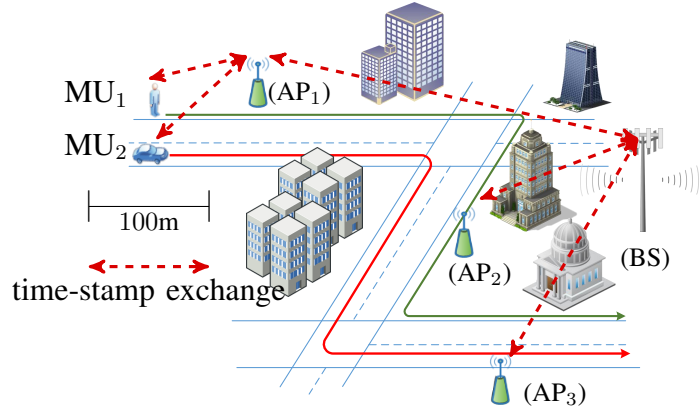
\begin{figure}[t!]
\begin{tikzpicture}[scale=1]
   \draw (0, 0) node[inner sep=0] {\includegraphics[width=0.48\linewidth]{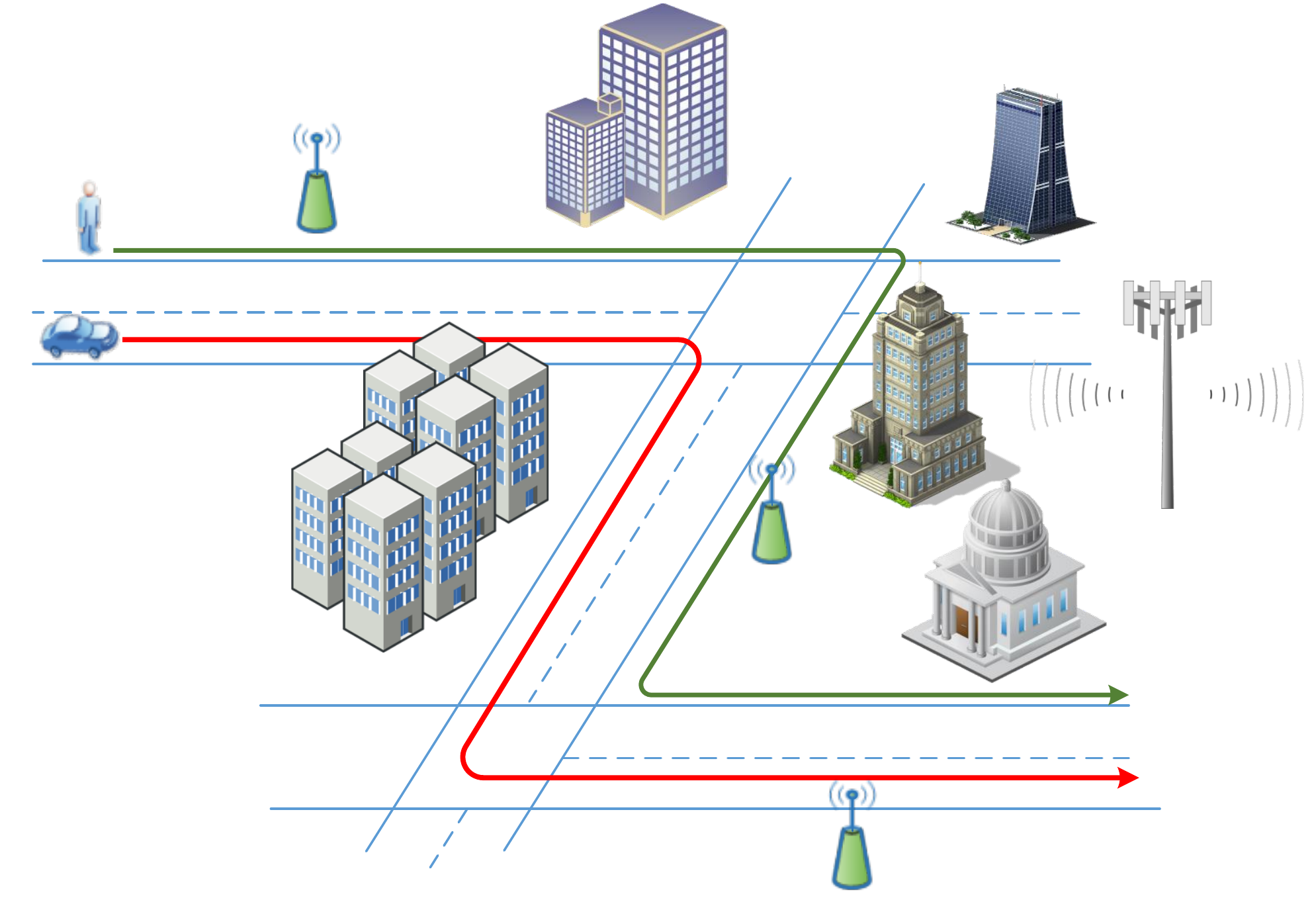}};
   \draw (-4.2,1.5) node(a0){MU$_1$};
   \draw (-4.2,0.8) node(a1){MU$_2$};
   \draw (-1.5,1.5) node(an1){\small(AP$_1$)};
   \draw (.9, -.9) node(an2){\small(AP$_2$)};
   \draw (1.9,-2.5) node(an3){\small(AP$_3$)};
   \draw (3.1, -.7) node(an2){\small(BS)};
   
   \draw (-4.3,0.) node(a0){$|$};
   \draw (-2.7,0.) node(a1){$|$};
   \draw (-3.5,-0.2) node(a2){\small $100$m};
   
   \draw[<->,red!85!black, very thick, dashed, >=stealth'] (-1.9,2.0) node (an1){} -- (3., 0.9) node (bs){};
   \draw[<->, red!85!black, very thick, dashed, >=stealth'] (.9,0.1) node (an1){} -- (3., 0.9) node (bs){};
   \draw[<->, red!85!black, very thick, dashed, >=stealth'] (1.3,-2.0) node (an1){} -- (3., 0.9) node (bs){};
   
   \draw[<->,red!85!black, very thick, dashed, >=stealth'] (-2.2,2.0) node (an1){} -- (-3.5, 1.7) node (bs){};
   \draw[<->,red!85!black, very thick, dashed, >=stealth'] (-2.2,2.0) node (an1){} -- (-3.4, .9) node (bs){};
   
   \draw[-] (a1.center) -- (a0.center);
   \draw[<->, red!85!black, very thick, dashed, >=stealth'] (-4.3,-.8) node (an1){} -- (-2.7, -.8) node (bs){} node[midway, below]{{\color{black}{time-stamp exchange}}};
\end{tikzpicture}
\centering
\caption{An example where MU joint sync\&loc can be carried out.}
\label{fig:scenario}
\end{figure}
The rest of this paper is structured as follows: In Section II, we introduce the system model and the preliminaries. Section III describes the details of the DePF algorithm for joint estimation of the clock and position parameters. Furthermore, the simulation results are presented and discussed in Section IV. Finally, Section V concludes this work and points to future works.
\subsubsection*{Notation} The boldface capital $\boldsymbol{A}$ and lower case $\boldsymbol{a}$ letters denote matrices and vectors, respectively. The $n$-th element of vector $\boldsymbol{a}$ is indicated by $\boldsymbol{a}[n]$. The symbol ``$\bullet$'' shows the inner scalar product of two (or multiple) vectors of the same dimension. Moreover, $\boldsymbol{I}_N$ and $\boldsymbol{0}_N$ represents $N\times N$ dimensional identity and all-zero matrices, respectively. $\mathbf{1}_N$ indicates an $N$-element all-one vector. Notation $\mathcal{U}(a, b)$ denotes a continuous uniform probability distribution in the interval between $a$ and $b$ with the probability level of $\frac{1}{b-a}$. Furthermore, $\mathcal{N}(\mathbf{x}|\boldsymbol{\mu}, \boldsymbol{\Sigma})$ stands for \textit{probability density function} (pdf) of a Gaussian random vector $\mathbf{x}$ with mean vector $\boldsymbol{\mu}$ and covariance matrix $\boldsymbol{\Sigma}.$ A diagonal matrix with the diagonal elements $(x_1, \cdots, x_K)$ is denoted by $\diag(x_1, \cdots, x_K)$. Symbol $\thicksim$ stands for ``is distributed as" and the symbol $\propto$ represents
the linear scalar relationship between two real-valued functions.
\section{System Model and Preliminaries}
We consider a network of multiple APs with known locations, all backhauled by BSs. The APs are assumed to feature multiple-input multiple-output Uniform Planar Arrays (UPAs), which allow for accurate azimuth and elevation AoA estimations. A further assumption is that they are able to continuously synchronize themselves with the backhauling BSs using the hybrid synchronization algorithm described in \cite{ goodarzi2020hybrid, goodarzi2021synchronization}. This, in particular, guarantees a low time error among the neighboring APs, enabling a more precise cooperative localization. Moreover, at each sync\&loc period $T$, a set of APs denoted by $\mathcal{I}_i,$ can periodically exchange time-stamps with the {$i$-th} MU using the FTM feature embedded in the communication devices and implemented by an existing protocol, e.g., precision time protocol \cite{eidson2002ieee}. From the packet containing these time-stamps, the APs can also estimate the CIRs and AoA. The \mbox{AP-MU} link condition is probabilistically determined and can be either LoS or NLoS. It is known from \cite{mondal20153d}, that, for such a scenario, the LoS probability is around 0.8, even growing to 0.95 when the AP density is 40 meters. A DNN trained using CIRs is employed to distinguish the LoS condition from NLoS, permitting the localization unit to neglect the measurements conducted under the NLoS condition, thereby augmenting the accuracy of synchronization and localization. In what follows, we firstly present the clock model for the APs and the MUs. Then, we explain the time-stamp exchange mechanism in detail. Subsequently, we discuss the DNN that allows for a reliable NLoS identification. Lastly, the principles of MUSIC algorithm are briefly described. 
\subsection{Clock Model}
We begin with defining a clock model for MUs and APs. For each node $i,$ we can write
\begin{equation}
c_i(t) = \gamma_i t + \theta_i,
\label{eq:clkmod}
\end{equation}
where $t$ represents the global reference time. Furthermore, $\gamma_i$ and $\theta_i$ denote the clock skew and the clock offset, respectively. Although the parameter $\gamma_i$ is generally random and time-varying, it is common to assume that it remains constant in the course of one synchronization period $T$ \cite{etzlinger2014cooperative,leng2011distributed,giorgi2011performance}. Given that, the first goal of the joint sync\&loc algorithm is to estimate and track the clock parameters $\gamma_i$ and $\theta_i$ (or transformations thereof) for each MU. In the sequel, we further clarify the components constructing $\theta_i$ as well as the time-stamp exchange mechanism required to estimate the above-mentioned parameters.
\begin{figure}[t!]
\includegraphics[width=0.6\linewidth]{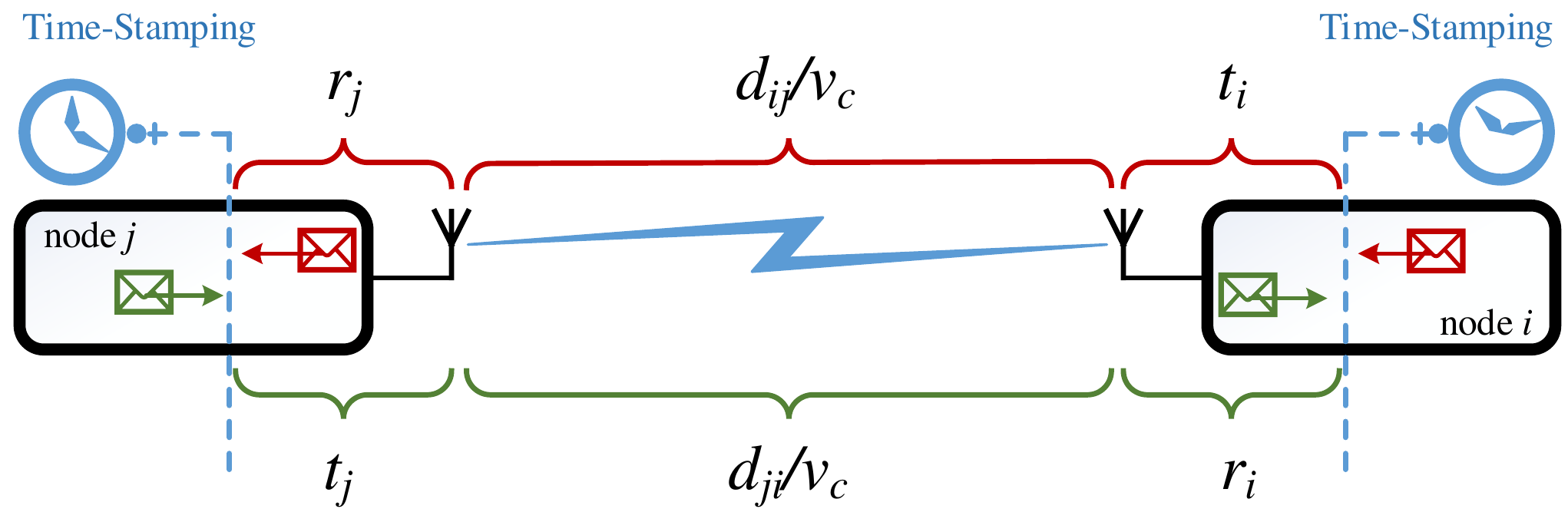}
\centering
\caption{Decomposition of the clock offset into its constituent components.}
\label{fig:deldec}
\end{figure}
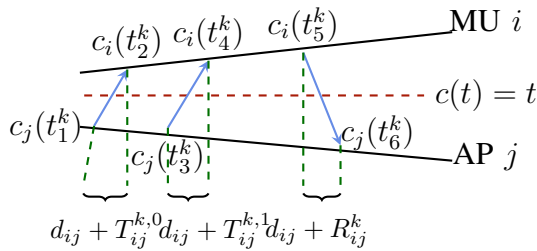
\begin{figure}[t!]
\begin{tikzpicture}[scale=.9]
	\definecolor{mc}{rgb}{0.4, 0.55, 0.9};
	\definecolor{mcshad}{rgb}{0.4, 0.55, 0.7};
	\definecolor{mc1}{rgb}{0.66, 0.11, 0.03};
	\definecolor{mc2}{rgb}{0.0, 0.44, 0.0};
	
	\draw (-3.5, 0.5) node(ci0){};
	\draw (-3.5, -0.3) node(cj0){};
	\draw (-3.5, -1.3) node(cj1){};
	\draw (2, 1.1) node(ci0end){};
	\draw (2.0, -0.8) node(cj0end){};
	\draw (2.0, -1.8) node(cj1end){};
	\draw[thick,-] (ci0.south)--(ci0end.south);
    \draw[thick,-] (cj0.south)--(cj0end.south);
    
	\draw (-3.6, 0) node(tbeg){};
	\draw (2.5, 0) node(tend){$c(t)=t$};
	\draw[mc1,thick,dashed] (tbeg.east)--(tend.west);
	
    \draw (-2.8, 0.8) node(cj2){$c_i(t_2^{k})$};
    \draw (-2.8, 0.8) node(cj2tex){};
    \draw (-0.2, 1.05) node(ci5) {$c_i(t_5^{k})$};
    \draw (-3.4, -0.8) node(ci1){};
    \draw (-4.0, -0.45) node(ci1){$c_j(t_1^{k})$ };
    \draw (0.35, -0.9) node(cj4){};
    \draw (0.9, -0.55) node(cj4tex){$c_j(t_6^{k})$};
    \draw (0, -0.85) node(cii){};
    \draw (-1.6, 0.93) node(ci4){$c_i(t_4^{k})$};
    \draw (-2.2, -0.9) node(cj3){$c_j(t_3^{k})$};
   
    \draw[mc,thick,-stealth] (cj3.north)--(ci4.south);
    \draw[mc,thick,-stealth] (ci1.east)--(cj2.south);
    \draw[thick, mc,-stealth] (ci5.south)--(cj4.north);
    
    \draw (-2.8, -1.35) node(cl2){};
    \draw (-1.6, -1.5) node(cl3){};
    
    \draw (-3.45,-1.5) node(b1){};
    \draw (-2.8,-1.5) node(b2){};
    \draw (-1.6,-1.5) node(b6){};
    \draw (-0.2,-1.5) node(e1){};
    \draw (0.35,-1.5) node(e2){};
    \draw (-2.2,-1.5) node(e3){};
    \draw (-3.1,-2.) node(b3){\small $d_{ij}+T_{ij}^{k, 0}$};
    \draw (-0,-2.) node(b4){\small $d_{ij}+R_{ij}^k$};
    \draw (-1.5,-2.) node(b5){\small $d_{ij}+T_{ij}^{k, 1}$};
    \draw[thick,decorate, decoration={brace,mirror,raise=0.1cm}] (b1.north) -- (b2.north);
    \draw[mc2,thick,dashed] (cj2.south)--(b2.north);
    \draw[mc2,thick,dashed] (ci1.east)--(b1.north);
    \draw[thick,decorate,decoration={brace,mirror,raise=0.1cm}] (e1.north) -- (e2.north);
    \draw[mc2,thick,dashed] (cj4.north)--(e2.north);
    \draw[mc2,thick,dashed] (ci5.south)--(e1.north);
    
    \draw[mc2,thick,dashed] (cj3.north)--(e3.north);
    \draw[mc2,thick,dashed] (ci4.south)--(b6.north);
    \draw[thick,decorate,decoration={brace,mirror,raise=0.1cm}] (e3.north) -- (b6.north);
    
    \draw (2.5, 1.1) node(ci5) {MU $i$};
    \draw (2.5, -.9) node(ci5) {AP $j$};
\end{tikzpicture}
\centering
\caption{Asymmetric time-stamp exchange between MU $i$ and AP $j$.}
\label{fig:stamp}
\end{figure}
\subsection{Offset Decomposition and Time-stamp Exchange}\label{ssec:offdec}
\subsubsection{Offset decomposition} To elaborate on the constituents of the offset $\theta_i$, we break it down as shown in Figure \ref{fig:deldec}. The parameter $t_j$/$t_i$ is the time taken for a packet to leave the transmitter after being \mbox{time-stamped}, $d_{ji}$/$d_{ij}$ represents the distance between the nodes $j$/$i$ and $i$/$j$, $v_c$ is the speed of light, and $r_i$/$r_j$ represents the time that a packet needs to reach the time-stamping point upon arrival at the receiver. Generally, the packets sent from node $j$ to node $i$ do not necessarily experience the same delay as those sent from node $i$ to node $j.$ In other words, $$t_j + \frac{d_{ji}}{v_c} + r_i \neq t_i + \frac{d_{ij}}{v_c} + r_j.$$ The variables $T_{ij} = t_j + r_i,$ and $R_{ij} = t_i + r_j$ (and correspondingly $t_j,$ $t_i,$ $r_j,$ and $r_i,$) are random variables due to multiple \mbox{hardware-related} random independent processes and can, therefore, be assumed i.i.d. Gaussian random variables, whereas $d_{ji}$ and $d_{ij}$ are usually assumed to be deterministic and symmetric ($d_{ji} = d_{ij}$) \cite{wu2010clock,leng2011distributed}.
The random variables $T_{ij}$ and $R_{ij}$ are assumed to be distributed as $\mathcal{N}(T_{ij}|\mu_T, \sigma^2_T)$ and $\mathcal{N}(R_{ij}|\mu_R, \sigma^2_R),$ respectively. As mentioned in \cite{leng2011distributed,du2013distributed,etzlinger2014cooperative}, while it is typical to assume that $\mu_T = \mu_R,$ and parameters $\sigma_{T}$ and $\sigma_{R}$ are known,
having any information about the value of $\mu_T$ and $\mu_R$ is highly unlikely. Therefore, we construct the joint sync\&loc algorithm assuming no knowledge on $\mu_T$ and $\mu_R$ except for $\mu_T = \mu_R.$
\subsubsection{Time-stamp exchange mechanism}\label{sssec:schedule}
We draw on the asymmetric time-stamp exchange mechanism shown in Figure \ref{fig:stamp}, proposed in \cite{chepuri2012joint}, and employed in \cite{etzlinger2014cooperative,etzlinger2017cooperative}. Node $j$ transmits a \textit{sync} message wherein the local time $c_j(t_1^k)$ is incorporated. Node $i$ receives the packet and records the local reception time $c_i(t_2^k)$. After a certain time, the process repeats again with $c_j(t_3^k)$ and $c_i(t_4^k).$ Subsequently, at local time $c_i(t_5^k)$, node $i$ sends back a \textit{sync} message to node $j$ with $c_i(t_2^k),$ $c_i(t_4^k)$ and $c_i(t_5^k)$ incorporated. Upon reception, node $j$ records the local time $c_j(t_6^k).$ Given this mechanism, at the $k$-th round of time-stamp exchange (and correspondingly $k$-th round of joint sync\&loc), we expect the localization unit to have collected the time-stamps  
$$\ccij^k = \left[c_j(t_1^k), c_i(t_2^k), c_j(t_3^k), c_i(t_4^k), c_i(t_5^k), c_j(t_6^k)\right]. $$
The collected time-stamps will be exploited in Section \ref{sec:paramest} to design a joint sync\&loc algorithm.
In the following subsections, we firstly use the CIRs to identify whether the MU-AP link condition is LoS or NLoS. Later on, the same CIRs are utilized to estimate the AoA.
\subsection{NLoS Identification and Channel Impulse Response}
The ability to estimate the CIR is highly ubiquitous among the APs. Therefore, relying on the CIR to develop a localization algorithm appears to be a realistic approach. The \mbox{AP-MU} CIR is a rich source of information about the condition of the communication link, e.g., LoS or NLoS, and the location of the MU. More precisely, the former is crucial to know when estimating the latter as the accuracy of the distance/time and AoA measurements significantly decline when conducted under NLoS conditions.

\begin{figure}[t!]
\includegraphics[width=0.55\linewidth]{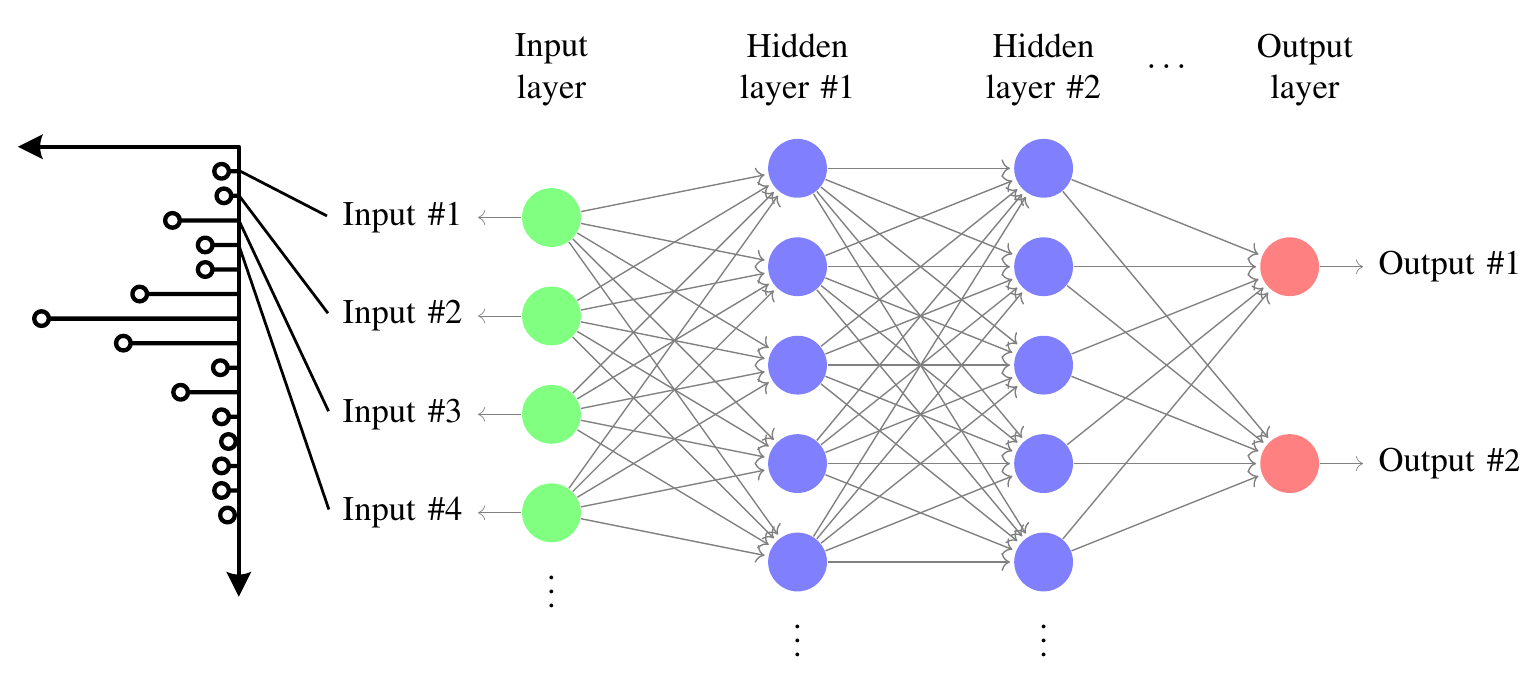}
\centering
\caption{The DNN employed for NLoS-identification. It has $l_H=2$ hidden layers with $n_H$ neurons and two output neurons.}
\label{fig:dnn}
\end{figure}
Figure \ref{fig:dnn} shows the architecture of the DNN deployed for NLoS-identification. The input layer has one channel fed with $N$ samples, i.e., the magnitude of the CIR. The number of hidden layers and neurons in each hidden layer is set to $l_H$ and $n_H$, respectively. The rationale to rely on when selecting these numbers is that, according to \cite{heaton2008introduction}, any classifier function can be realized by two hidden layers, i.e., currently there is no theoretical reason to use more than two. However, the lack of evidence does not imply that the DNNs with more hidden layers do not improve the accuracy of classification, it rather suggests that the number of required hidden layers does not follow a {well-established} logic and is mostly determined by a trial-and-error process. Therefore, for the algorithm proposed in this work, we empirically determine the $l_H$ that delivers the best performance. Furthermore, as a rule of thumb, the number of neurons is suggested to be between the number of inputs and that of the outputs to prevent under/overfitting.

Let the output probability vector of the DNN be $[1-\cn, \cn],$
where $\cn$ denotes the probability of the CIR being corresponded to an NLoS link. For the NLoS-identifier, we seek to train the DNN such that the output probability vector is as close as possible to $[1, 0]$/$[0, 1]$ for the LoS/NLoS CIRs. In other words, from the optimization point of view, we aim to design a loss function whose output is small when the DNN returns the correct vector and it is large otherwise. It turns out that the function that possesses the above-mentioned property is the logarithmic function \cite{boyd2004convex}. Mathematically, the loss function is given by \cite{liu2017survey}
\begin{align}
\mathcal{L} & = -\frac{1}{M_c}\sum_{i=1}^{M_c} p_{\text{nlos}}^i\log(\hat{p}_{\text{nlos}}^i) + (1-p_{\text{nlos}}^i)\log(1-\hat{p}_{\text{nlos}}^i),
\label{eq:bce}
\end{align}
where $p_{\text{nlos}}^i$ denotes the true label corresponding to the $i$-th CIR sample in the data set and is one if the CIR corresponds to an NLoS link and zero otherwise. Furthermore, $M_c$ represents the total number of CIRs in the training set. The formulation in (\ref{eq:bce}) is also known in the literature as the binary cross-entropy loss function. The goal of training is then to adjust the weights of the neurons such that (\ref{eq:bce}) is minimized. Finally, when the trained DNN is employed in the context of joint sync\&loc algorithm, the decision on the link condition is fed into the algorithm using the binary parameter $\zeta_i,$ which is set to one when $\cn>0.5$ and zero otherwise. Specifically, if $\zeta_i$ is zero, the communication link is considered NLoS and any measurement corresponding to it, i.e., time-stamp exchange and AoA, is dropped.

In the sequel, we present the principles of the AoA estimation algorithm, which draws on the CIRs employed for NLoS identification
\subsection{Angle of Arrival}\label{sssec:crb}
The CIR fed into the DNN to identify the link condition can be treated as an input signal to the MUSIC algorithm to obtain the AoA. We present the principles of AoA estimation for UPAs based on \cite{chen2018two, yan2015two, mohanna2013optimization}. The estimated AoA is given by 
\begin{equation}
(\varphi_{ij},\alpha_{ij})  = \argmax_{\varphi, \alpha}\frac{1}{\mathbf{a}_n(\varphi, \alpha)^H\mathbf{N}\mathbf{N}^H\mathbf{a}_n(\varphi, \alpha)},
\label{eq:fi}
\end{equation}
where $\varphi_{ij}$ and $\alpha_{ij}$ are the azimuth and elevation AoA of the signal received from the MU $i$ at AP $j,$ respectively. Parameter $\mathbf{a}_n(\varphi, \alpha)$ is the signal vector rotation on the $n$-th subcarrier and is given by
\begin{multline}
\mathbf{a}_n(\varphi, \alpha) = \left[ 1,\  
 e^{i\frac{2\pi d}{\lambda}\sin(\alpha)\left(\sin(\varphi)+\cos(\varphi)\right)},\ 
 e^{i\frac{2\pi d}{\lambda}\sin(\alpha)\left(\sin(\varphi)+ 2\cos(\varphi)\right)},\ 
  \cdots, \right. \\
 \left. e^{i\frac{2\pi d}{\lambda}\sin(\alpha)\left((\nant-1)\sin(\varphi)+ (\nant -2 )\cos(\varphi)\right)},\ 
 e^{i\frac{2\pi d}{\lambda}(\nant-1)\sin(\alpha)\left(\sin(\varphi)+\cos(\varphi)\right)} \right]^T_{1\times\nant^2}
\end{multline}
where $\nant$ denotes the number of AP antennas in one row (or column). Matrix $\mathbf{N}$ is constructed by $\nant^2 - 1$ most right columns of the eigenvectors obtained when performing the eigen decomposition of the covariance matrix of the received signal. That is 
\begin{equation}
\mathbf{R} = \mathbf{V} \mathbf{A} \mathbf{V}^H,
\label{eq:eigen}
\end{equation}
where matrices $\mathbf{A}$ and $\mathbf{V}$ contain the eigenvalues and eigenvectors, respectively. Furthermore,
\begin{equation}
\mathbf{R} = \frac{1}{N_s}\sum_{n=1}^{N_s} \mathbf{x}_n\mathbf{x}_n^H,
\end{equation}
where the vector $\mathbf{x}_n$ is of dimension $\nant^2\times 1$ and represents the $n$-th element of FFT of the CIRs. The number of subcarriers, or, alternatively, the size of FFT is denoted by $N_s$. It is worth mentioning that, when constructing $\mathbf{N},$ the eigen decomposition in (\ref{eq:eigen}) is assumed to sort the eigenvalues in decreasing order. Lastly, each AP is assumed to have $\nant^2$ CIRs at its disposal.
\section{Clock Parameters and Position Estimation}\label{sec:paramest}
In this section, we discuss an estimation method for the clock and position parameters. It relies primarily on the components analyzed in the previous section, i.e, time-stamp exchange, AoA estimation, and NLoS identification. In particular, given Section~\ref{sssec:schedule}, and considering AP $j$ as the master node, we can write
\begin{align}
&\tgam(c_i(t_{2}^k) - \tilde{\theta}_i) = c_j(t_1^k) + \frac{d_{ij}}{v_c} + \tij{k, 0}  ,\label{eq:c1}\\ 
&\tgam(c_i(t_{4}^k) - \tilde{\theta}_i) = c_j(t_3^k) + \frac{d_{ij}}{v_c} + \tij{k, 1}, \label{eq:c2}\\
&\tgam(c_i(t_{5}^k) - \tilde{\theta}_i) = c_j(t_6^k) - \frac{d_{ij}}{v_c} - \rij{k}, \label{eq:c3} 
\end{align}
where $t_1^k$/$t_2^k$, $t_3^k$/$t_4^k$, and $t_5^k$/$t_6^k$ are the time points where MU $i$ and AP $j$ send/receive the sync messages, respectively. Parameter $d_{ij} = \dij{j}{}$ denotes the Euclidean distance between nodes $i$ and $j$. We note that, in Figure \ref{fig:stamp}, instead of a global time reference $c(t)=t,$ we take node $j$ as the master node. It is straightforward to see that $\frac{1}{\tilde{\gamma}_i}=\frac{\gamma_j}{\gamma_i},$ $\tilde{\theta}_i = \theta_i-\tg\theta_j,$ $\tilde{d}_{ij} + \tijt{k} = \gamma_j(d_{ij} + \tij{k}),$ and $\tilde{d}_{ij} - \rijt{k} = \gamma_j(d_{ij} - \rij{k})$. For the sake of simplicity, as done in \cite{wu2010clock}, we assume $\tilde{d}_{ij}=d_{ij},$ $\rijt{k} = \rij{k},$ and $\tijt{k}=\tij{k}.$ This is valid because $\gamma_j\approx 1$ and the values of $d_{ij} + \tij{k}$ and $d_{ij} - \rij{k}$ are small. In what follows, we first give the probabilistic representation of the problem. Subsequently, the principles of the estimation method are presented.  
\subsection{Probabilistic Formulation of the Problem}
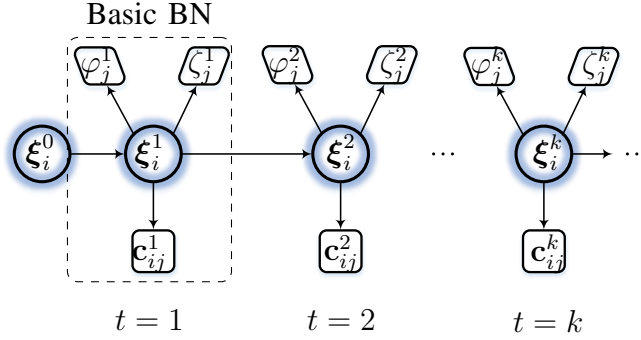
\begin{figure}[t!]
\begin{tikzpicture}[scale=1]
    \draw (0, 0) node[inner sep=0] {\includegraphics[width=.53\linewidth]{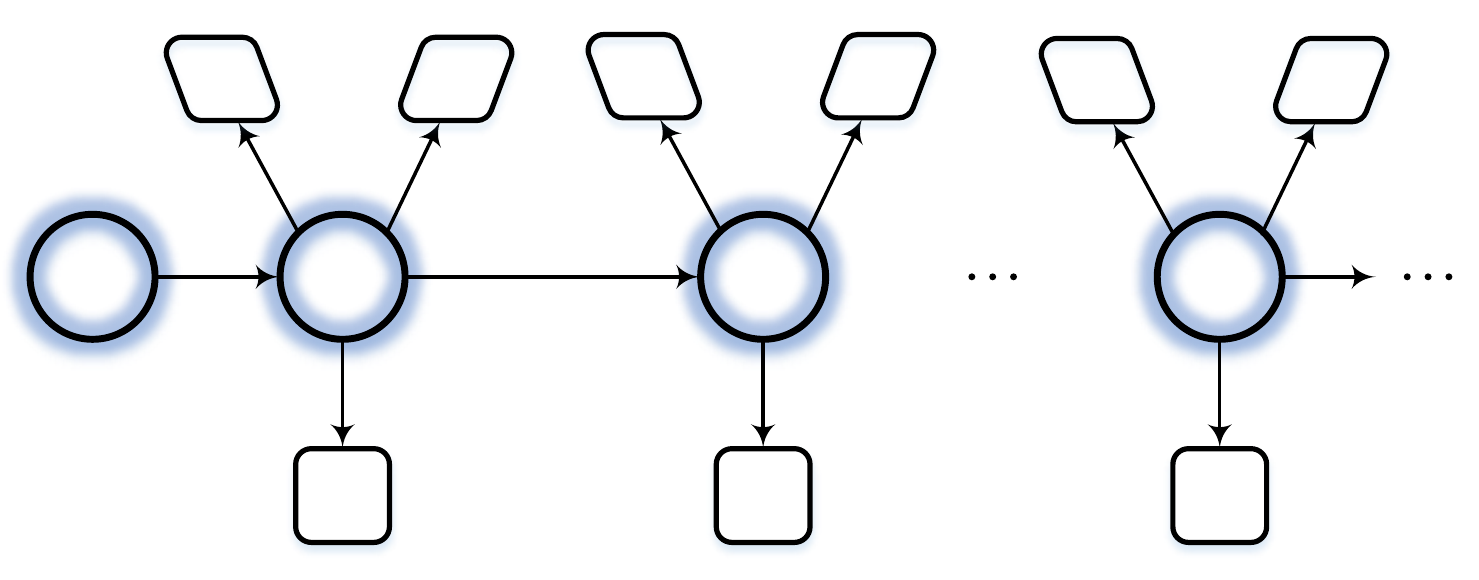}};
	\draw (-3.85,0.1) node(a00){$\tetvect{i}^0$};  
	\draw (-2.4,0.1) node(a00){$\tetvect{i}^1$};
	\draw (.15,0.1) node(a00){$\tetvect{i}^2$};
	\draw (2.9,0.1) node(a00){$\tetvect{i}^k$};
	
	\draw (-2.4,-1.25) node(c1){$\ccij^1$};
	\draw (.15,-1.25) node(c2){$\ccij^2$};
	\draw (2.9,-1.25) node(ck){$\ccij^k$};
	
	\draw (-3.1,1.25) node(a00){$\varphi_j^1$};
	\draw (-.6,1.25) node(a00){$\varphi_j^2$};
	\draw (2.1,1.25) node(a00){$\varphi_j^k$};
	
	\draw (-1.7,1.25) node(a00){$\zeta_j^1$};
	\draw (0.85,1.25) node(a00){$\zeta_j^2$};
	\draw (3.55,1.25) node(a00){$\zeta_j^k$};
	
	\draw (-2.4, 0) node[rectangle, dashed, draw, rounded corners, minimum height=3.25cm, text width=1.9cm]{};
	\draw (-2.4, 1.95)node(a00){Basic BN};
	
	\node [below= 0.2cm of c1]{$t=1$};
	\node [below= 0.2cm of c2]{$t=2$};
	\node [below= 0.2cm of ck]{$t=k$};
\end{tikzpicture}
\centering
\caption{Dynamic Bayesian network representing the temporal evolution of the vector variable $\ttat{}$ and its relation to the measurements.}
\label{fig:bayesrep}
\end{figure}

Let $\ttat{k}$ be the state of the vector variable $\tetvect{i} \triangleq \begin{bmatrix}\tetvec{i} & \pos \end{bmatrix}^T$ after the $k$-th round of time-stamp exchange, where $\tetvec{i} = \begin{bmatrix}\tgam  & \frac{\ttet}{\tg}\end{bmatrix}$ and  $\pos = \begin{bmatrix}x_i & y_i\end{bmatrix}.$ Parameters $x_i$ and $y_i$ denote the position of node $i$ on the $x$ and $y$ axes, respectively.  
The aim is then to infer the pdf corresponding to the $k$-th state, which can be written as
\begin{align}
 &p(\ttat{k}|\{\ccij^{1:k}, \varphi_{ij}^{1:k}, \zeta_{ij}^{1:k}\}_{\forall j\in \mathcal{I}_i}) = \int p(\ttat{0},\cdots, \ttat{k}|\{\ccij^{1:k}, \varphi_{ij}^{1:k}, \zeta_{ij}^{1:k}\}_{\forall j\in \mathcal{I}_i})\ d\ttat{0}\cdots d\ttat{k-1},
 \label{eq:pdfk}
\end{align}
where the superscript $1:k$ indicates the collection of measurements from the first round until the $k$-th. Applying Bayes rule, we can rewrite (\ref{eq:pdfk}) as
\begin{align}
&p(\ttat{k}|\{\ccij^{1:k}, \varphi_{ij}^{1:k}, \zeta_{ij}^{1:k}\}_{\forall j\in \mathcal{I}_i}) \propto \int p(\{\ccij^{1:k}, \varphi_{ij}^{1:k}, \zeta_{ij}^{1:k}\}_{\forall j\in \mathcal{I}_i}|\ttat{0},\cdots, \ttat{k})p(\ttat{0},\cdots, \ttat{k}) d\ttat{0}\cdots d\ttat{k-1}.
\label{eq:bayesrule}
\end{align}
Figure \ref{fig:bayesrep} depicts the temporal evolution of $\ttat{k}$ as well as its relation with the measurements at each time step. Such a structure is referred to as dynamic Bayesian Network (BN), in which a basic BN repeats itself in each time step \cite{mihajlovic2001dynamic}. The states of a dynamic BN, i.e., all the variables with the same time index, satisfy the Markov property\footnote{It postulates that the state of the system at time $t$ depends only on its immediate past, i.e. its state at time $t-1.$} \cite{barker1995bayesian}, enabling us to carry out the following mathematical simplifications. In particular,
knowing that the measurements are independent and assuming the Markov property, we reformulate the integrands in (\ref{eq:bayesrule}) as
\begin{align}
&p(\{\ccij^{1:k}, \varphi_{ij}^{1:k}, \zeta_{ij}^{1:k}\}_{\forall j\in \mathcal{I}_i}|\ttat{0},\cdots, \ttat{k}) =
 p(\{\ccij^{k}, \varphi_{ij}^{k}, \zeta_{ij}^{k}\}_{\forall j\in \mathcal{I}_i}|\ttat{k})\cdots p(\{\ccij^{1}, \varphi_{ij}^{1}, \zeta_{ij}^{1}\}_{\forall j\in \mathcal{I}_i}|\ttat{1}), \nonumber \\
&p(\ttat{0},\cdots, \ttat{k}) = \tntt{k}{k-1}\cdots \tntt{1}{0}p(\tetvect{i}^0),  
\label{eq:markov}
\end{align}
where $p(\tetvect{i}^0)$ denotes the prior knowledge on $\tetvect{i}.$ Plugging (\ref{eq:markov}) into (\ref{eq:bayesrule}) leads to
\begin{multline}
p(\ttat{k}|\ccij^{1:k}, \varphi_{ij}^{1:k}, \zeta_{ij}^{1:k}) \propto \\ \underbrace{\int p(\ttat{0})\left[\prod_{r=1}^{k-1}p(\ttat{r}|\ttat{r-1})\cntnt{r}{r}\right]\tnt{k}{k-1}}_{= p(\tta{k}|\ccij^{1:k-1}, \varphi_{ij}^{1:k-1}, \zeta_{ij}^{1:k-1})} \times \cntnt{k}{k} d\ttat{0}\cdots d\ttat{k-1}.
\label{eq:longeq}
\end{multline}
Finally, we can write
\begin{align}
&p(\ttat{k}|\{\ccij^{1:k}, \varphi_{ij}^{1:k}, \zeta_{ij}^{1:k}\}_{\forall j\in \mathcal{I}_i}) \propto p(\tetvect{i}^{k}|\{\ccij^{1:k-1}, \varphi_{ij}^{1:k-1}, \zeta_{ij}^{1:k-1}\}_{\forall j\in \mathcal{I}_i})\cntnt{k}{k}.
\label{eq:bayesfin}
\end{align}  
The term $p(\tetvect{i}^{k}|\{\ccij^{1:k-1}, \varphi_{ij}^{1:k-1}, \zeta_{ij}^{1:k-1}\}_{\forall j\in \mathcal{I}_i})$ is referred to as \textit{prediction} step while the term $\cntnt{k}{k}$ is considered as \textit{correction} step \cite{barker1995bayesian}. If the Gaussian assumption about $\tetvect{i}^0$ held and the relation between all the states in Figure \ref{fig:bayesrep} were linear, we could conclude that the marginal in (\ref{eq:bayesfin}) would also be Gaussian distributed. Unfortunately, that is not the case in the joint sync\&loc problem as the measurement equations (and consequently the correction steps) are partially \mbox{non-linear}. In concrete terms, the aforementioned problem stems from the nonlinear relation between the location parameters $(x_i, y_i)$ and the time-stamps in (\ref{eq:c1}), (\ref{eq:c2}), (\ref{eq:c3}) on one hand, and the measured AoA in (\ref{eq:fi}) on the other hand.

There are several approaches to tackle the nonlinearity problem and, consequently, to estimate the non-Gaussian posterior distribution. In \cite{goodarzi2020bayesian}, it is proposed to undertake the Taylor expansion of the nonlinear terms around the prediction point, while \cite{werner2015joint, koivisto2017joint, khan2014localization} have employed EKF to address the \mbox{non-linearity}. In addition to being prone to divergence, which is hard to mitigate analytically, all of these methods require initialization and even then are only able to deliver medium accuracy. In what follows, we discuss the details of a novel joint sync\&loc approach based on PGM filters.
\subsection{Particle Gaussian Mixure Filter}
The idea underpinning PGM filters is to approximate a pdf by the sum of weighted \textit{Gaussian density functions} (gdfs) \cite{alspach1972nonlinear}. Leveraging this idea, we can write the posterior in (\ref{eq:bayesfin}) as 
\begin{equation}
p(\tetvect{i}^{k}|\{\ccij^{1:k}, \varphi_{ij}^{1:k}, \zeta_{ij}^{1:k}\}_{\forall j\in \mathcal{I}_i}) = \sum_{f=1}^F w_f^k\mathcal{N}(\tetvect{i}^k|\boldsymbol{\mu}_f^k, \boldsymbol{\Sigma}_f^k),\ \ \text{with}\ \  \sum_{f=1}^F w_f^k = 1, \ \ w_f^k\geq 0\ \forall f,
\label{eq:pgm}
\end{equation}
where $\boldsymbol{\mu}_f^k = \begin{bmatrix}
\mugg{\tetvec{i}} & \mugg{\pos}
\end{bmatrix}$ and $\boldsymbol{\Sigma}_f^k = \begin{bmatrix}
\sigg{\tetvec{i}}_f^k & \mathbf{0}_2 \\ \mathbf{0}_2 & \sigg{\pos}_f^k \end{bmatrix}$ denote the mean vector and covariance matrix of the $f$-th gdf in the $k$-th round of estimation, respectively. Parameter $F$ represents the total number of gdfs. Furthermore, $\mugg{\tetvec{i}}$/$\mugg{\pos}$ and $\sigg{\tetvec{i}}_f^k$/$\sigg{\pos}_f^k$ represent the mean vector and covariance matrix corresponding to the vector variable $\tetvec{i}$/$\pos$, respectively.

\begin{figure}[t!]
\begin{tikzpicture}[scale=1]
\draw (0, 0) node[inner sep=0] {\includegraphics[width=.55\linewidth]{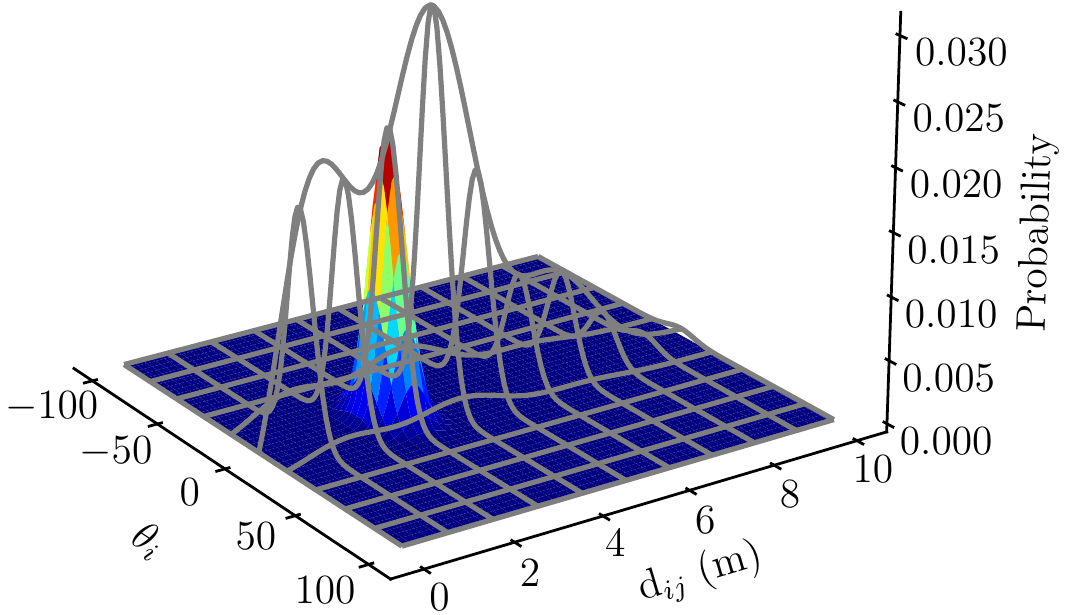}};
\node (A) at (-2., 3) {A single Gaussian mixture (gdf)};
 \node (B) at (-1.3, 1.3) {};
 \draw[->, black, thick, >=stealth'] (A) --  (B){};

\node (A) at (1.3, 1.8) {non-Gaussian pdf};
 \node (B) at (-.6, 1.3) {};
\draw[->, black, thick, >=stealth'] (A) --  (B){};
V
\end{tikzpicture}
\centering
\caption{An example distribution of the $\tetvect{i}$ for a given  time-stamp measurement.}
\label{fig:exdist}
\end{figure}
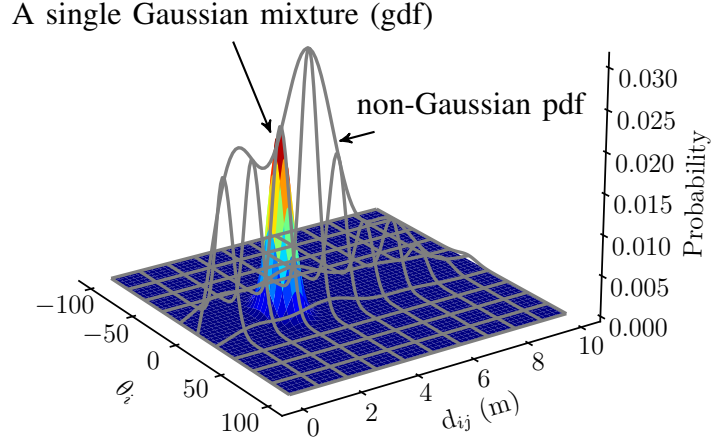

Seeking to further simplify (\ref{eq:pgm}), we reformulate (\ref{eq:c1}), (\ref{eq:c2}), and (\ref{eq:c3}) as follows. Subtracting (\ref{eq:c1}) from (\ref{eq:c2}) leads to
\begin{align}
& \frac{1}{\tilde{\gamma}_i}(c_i(t_{4}^{k}) - c_i(t_{2}^{k})) = c_j(t_3^{k}) - c_j(t_1^{k}) + \tij{k, 1}-\tij{k, 0}, \label{eq:c2-c1}
\end{align} 
while summing up (\ref{eq:c2}) and (\ref{eq:c3}) gives
\begin{align}
&\tgam(c_i(t_{4}^k) + c_i(t_{5}^k)-2\ttet) = c_j(t_3^k) + c_j(t_6 ^k) + \tij{k, 1}-\rij{k}. \label{eq:c2+c3} 
\end{align}
It is straightforward to observe that $\tetvec{i}^k,$ on one hand, is linearly dependent on the time-stamps, and, on the other hand, does not depend on $\pos.$ This suggests that, although the $\cntnt{k}{k}$  is not Gaussian distributed in general, it is indeed Gaussian across the $\tetvec{i}$ axis as both $\tij{}$ and $\rij{}$ are Gaussian distributed. We capitalize on the linear Gaussian substructures in the model to keep the state dimensions low.  Consequently, the gdfs can be employed only across the $\pos$ axis transforming the structure of (\ref{eq:pgm}) into the multiplication of a single gdf across $\tetvec{i}$ and sum weighted of multiple gdfs across $\pos$ (visualized in Figure \ref{fig:exdist}). Such a structure not only lays the ground for the hybrid parametric and particle-based implementation of \mbox{BRF-based} joint sync\&loc estimation but also dramatically reduces the computational burden.
Given above, (\ref{eq:pgm}) can be simplified as
\begin{align}
&p(\{\ccij^{1:k}, \varphi_{ij}^{1:k}, \zeta_{ij}^{1:k}\}_{\forall j\in \mathcal{I}_i}|\tetvect{i}^{k}) =  \mathcal{N}(\tetvec{i}^k|\mug{\tetvec{i}}^k, \sigg{\tetvec{i}}^k)\sum_{f=1}^F w_f^k\mathcal{N}(\mathbf{p}_i^k|\mugg{\pos}, \sigg{\pos}_f^k).
\label{eq:pgmf}
\end{align}
We note that when $\sigg{\pos}_f^k$ approaches $0,$ the term $\mathcal{N}(\pos^k|\mugg{\pos}, \sigg{\pos}_f^k)$ tends towards $\delta(\pos^k-\mugg{\pos}),$ where $\delta(\cdot)$ denote the Dirac impulse function. Such a function forms the basis of the classical particle filter. In what follows, we further delve into the steps of parameter estimation of the above-mentioned distribution. Firstly, the details of \textit{prediction} step are described, where all the parameters are denoted by $(\cdot)_-.$ Next, we obtain the likelihood of the measurements whose parameters are represented by $(\cdot)_+.$ Lastly, we compute the parameters of the posterior distribution in (\ref{eq:pgmf}) and perform the resampling.
\subsubsection{Prediction} Given the linear dynamics of MUs' clocks and movements, a reasonable prediction for $\tetvect{i}^k$ is given by 
\begin{align}
&p(\tetvect{i}^{k}|\{\ccij^{1:k-1}, \varphi_{ij}^{1:k-1}, \zeta_{ij}^{1:k-1}\}_{\forall j\in \mathcal{I}_i}) =  \mathcal{N}(\tetvec{i}^k|\mug{\tetvec{i}}^k_-, \sigg{\tetvec{i}}^k_-)\sum_{f=1}^F w_{f-}^k\mathcal{N}(\mathbf{p}_i^k|\mug{\pos}^k_{f-}, \sigg{\pos}^k_{f-})
\label{eq:predpdf}
\end{align} 
where
\begin{align*}
&w_{f-}^k = \frac{1}{F}\mathbf{1}_F, & \mug{\pos}^k_{f-} = \mug{\pos}^{k-1}_f + \mathbf{n}_f,& 
\end{align*}
with $\mathbf{n}_f$ being the noise vector derived from the distribution $\mathcal{N}(\mathbf{n}|\mathbf{0}, \qn{\pos}),$ for $\qn{\pos}=\diag(\sigma^2_x,\sigma^2_y).$ In practice, we initialize $\sigg{\pos}^k_{f-} \propto \diag(F^{-0.4}, F^{-0.4}),$ which is proved in \cite{stordal2011bridging} to be the optimal choice. Furthermore, according to \cite{goodarzi2020hybrid},
\begin{align}
&\mug{\tetvec{i}}^k_- = \mathbf{F}\mug{\tetvec{i}}^{k-1} + \mathbf{u},& \sigg{\tetvec{i}}^k_- = \mathbf{F}\sigg{\tetvec{i}}^{k-1}\mathbf{F}^T + \qn{\tetvec{i}} &
\label{eq:predpdfxy}
\end{align}
with \begin{align*}
& \mathbf{F}=\begin{bmatrix} 1 & 0 \\   T & 1 \end{bmatrix}, & \mathbf{u}=\begin{bmatrix} 0 & T  \end{bmatrix}^T, &\  \qn{\tetvec{i}} = \diag(\sigma_{\gamma}^2, \sigma_{\theta}^2).
\end{align*}
The matrices $\qn{\tetvec{i}}$ and $\qn{\pos}$ denote the covariance of the zero-mean Gaussian noises on each gdf across the $\tetvec{i}$ and $\pos$ axes, respectively. In general, the design of $\qn{\cdot}$ is a difficult task. In particular, if it is too small, the filter will be overconfident in its prediction model and will diverge from the actual solution. In contrast, if it is too large, it will be unduly dominated by the noise in the measurements and perform sub-optimally \cite{labbe2019kalman}. Similar to \cite{etzlinger2014cooperative, wu2010clock, giorgi2011performance}, we set $\sigma_{\gamma}^2$ and $\sigma_{\theta}^2,$ such that the external noises as well as the residues from the previous iteration are accounted for. Furthermore, to determine the value of $\sigma^2_x$ and $\sigma^2_y,$ the design model discussed in \cite{labbe2019kalman,khan2014localization} is followed. That is, opting for a noise variance that is large enough to allow the gdfs to assign a reasonable probability to the locations where the MU might be. In the urban scenario, for example, the maximum permitted speed is 50 km/h ($\approx$14 m/s), resulting in $\sigma_x = \sigma_y = 14\times T.$ 
\subsubsection{Measurement Likelihood and Weight Update}
The same structure as (\ref{eq:pgmf}) is imposed on the likelihood of the measurements. That is,
\begin{align}
&\cntnt{k}{k} =  \mathcal{N}(\tetvec{i}^k|\mug{\tetvec{i}}^k_+, \sigg{\tetvec{i}}^k_+)\sum_{f=1}^F w_{f+}^k\mathcal{N}(\mathbf{p}_i^k|\mug{\pos}^k_{f+}, \sigg{\pos}^k_{f+}).
\label{eq:mespdf}
\end{align} 
To obtain the parameters of the above likelihood, we firstly transform (\ref{eq:c2-c1}) and (\ref{eq:c2+c3}) into the matrix form. That is,   
\begin{equation}
\mathbf{B}_{ij}^k\tetvec{i}^k = \mathbf{r}_{ij}^k + \zji,
\label{eq:meas_eq}
\end{equation}
where $\zji\sim \mathcal{N}(\mathbf{z}|\mathbf{0},\mathbf{R}_{ij}^k)$ with $\mathbf{R}_{ij}^k = \diag(2\sigma^2_{T_{ij}}, \sigma^2_{T_{ij}}+\sigma^2_{R_{ij}}),$ and
 $$ \mathbf{B}_{ij}^k = \begin{bmatrix}
  c_i(t_{4}^k)-c_i(t_{2}^{k}) & 0 \\
 c_i(t_{4}^k)+c_i(t_{5}^k) & -2 
 \end{bmatrix}, \mathbf{r}_{ij}^k = \begin{bmatrix} c_j(t_3^k)-c_j(t_1^k) \\ c_j(t_3^k)+ c_j(t_6^k)
\end{bmatrix}.$$
The mean and covariance matrix of the gdfs across the $\tetvec{i}$ axis can be written as
\begin{align}
&\mug{\tetvec{i}}^k_+ = \mathbf{A}_{ij}^k\mathbf{r}_{ij}^k, & \sigg{\tetvec{i}}^k_+ = \mathbf{A}_{ij}^k\mathbf{R}_{ij}^k (\mathbf{A}_{ij}^k)^T,&
\label{eq:musigzz}
\end{align}
where $\mathbf{A}_{ij}^k = ((\mathbf{B}_{ij}^k)^T\mathbf{B}_{ij}^k)^{-1}(\mathbf{B}_{ij}^k)^T.$ 

To obtain the location parameters corresponding to each gdf, we can assume that the measurement equations are linear in the vicinity of the point predicted by the prediction step. That is, to approximate them with their first-order Taylor expansions, the details of which are thoroughly explained in \cite{goodarzi2020bayesian, goodarzi2021synchronization}\footnote{Note that this is equivalent to EKF, e.g. that of \cite{khan2014localization}. Nevertheless, to keep consistency with the approach taken in this work, i.e., Bayesian representation of the filtering process, we avoid the EKF representation.}. The measurement equations we rely on to estimate the parameters of the likelihoods are (\ref{eq:c3}) and
\begin{align}
\arctan(\frac{y_i-y_j}{x_i-x_j}) = \varphi_{ij}^k,\label{eq:aoa} 
\end{align} 
where $\varphi_{ij}^k$ is calculated as explained in Section \ref{sssec:crb}. Carrying out the necessary mathematical manipulation, we can write the same relation as (\ref{eq:meas_eq}) for each gdf. That is, 
\begin{equation}
\mathbf{B}_{ij, f}^k\pos^k = \mathbf{r}_{ij, f}^k + \mathbf{z}_{ij, f},
\label{eq:meas_eq_mix}
\end{equation}
where $\mathbf{z}_{ij, f}\sim \mathcal{N}(\mathbf{z}|\mathbf{0},\mathbf{R}_{ij, f})$ with $\mathbf{R}_{ij, f} = \diag(\sigma^2_{R_{ij}}, \sigma^2_{\varphi}).$ Furthermore,
  $\mathbf{B}_{ij, f} = \begin{bmatrix} 
 \mathbf{a}_j^k  & \mathbf{b}_j^k
\end{bmatrix}^T$
with the vectors $\mathbf{a}_j^k$ and $\mathbf{b}_j^k$ calculated by means of (\ref{eq:constay1}) and (\ref{eq:constay2}), respectively. Finally, $\mathbf{r}_{ij, f}$ is constructed as in (\ref{eq:rij}). 
\begin{figure*}
\begin{align}
&a^k_{j, f} = \frac{1}{v_c}\left|\mug{\pos}_{f-}^k-\mathbf{p}_j\right|,  && \mathbf{a}_{j, f}^{k} = \frac{1}{v_c^2 a^k_{j, f}}\left(\mug{\pos}_{f-}^k - \mathbf{p}_j\right), \label{eq:constay1}\\ 
&b^k_{j, f} = \arctan(\frac{\mathbf{a}_{j, f}^{k}[2]}{\mathbf{a}_{j, f}^{k}[1]}), && \mathbf{b}_{j, f}^k = \frac{1}{a^k_{j, f}}\begin{bmatrix}
-\mathbf{a}_{j, f}^{k}[2], & \mathbf{a}_{j, f}^{k}[1] \end{bmatrix}.\label{eq:constay2}
\end{align}
\begin{align}
&&\mathbf{r}_{ij, f} = \begin{bmatrix}c_j(t_{6}^k) - a_{j,f}^k +\mug{\pos}^k_{f-}\bullet \mathbf{a}_{j, f}^k - \begin{bmatrix}
c_i(t_{5}^k) & -1 \end{bmatrix}\bullet \mug{\tetvec{i}}^k_{+}, & \varphi_{ij}^k - b_{j, f}^k + \mug{\pos}^k_{f-}\bullet \mathbf{b}_{j, f}^k
\end{bmatrix}^T.&
\label{eq:rij}
\end{align}
\hrule
\end{figure*}
We note that (\ref{eq:constay1}) and (\ref{eq:constay2}) are computed by means of the Taylor expansion of (\ref{eq:c3}) and (\ref{eq:aoa}) around the predicted point $\mug{\pos}^k_{f-}$ with the known $\mug{\tetvec{i}}^k_+$ obtained by (\ref{eq:musigzz}). Given (\ref{eq:meas_eq_mix}), and similar to (\ref{eq:musigzz}), we can write 
\begin{align}
&\mug{\pos}^k_{f+} = \mathbf{A}_{ij, f}^k\mathbf{r}_{ij, f}^k, & \sigg{\pos}^k_{f+} = \mathbf{A}_{ij, f}^k\mathbf{R}_{ij, f}^k (\mathbf{A}_{ij, f}^k)^T,&
\label{eq:musigz}
\end{align}
where $\mathbf{A}_{ij, f}^k = ((\mathbf{B}_{ij, f}^k)^T\mathbf{B}_{ij, f}^k)^{-1}(\mathbf{B}_{ij, f}^k)^T.$ Furthermore, it is straightforward to see that 
\begin{equation}
w^k_{f+} = \mathcal{N}(\pos^k = \mug{\pos}^k_{f+}|\mug{\pos}^k_{f+}, \sigg{\pos}^k_{f+}).
\end{equation}
In other words, the weights are equal to the likelihood of the mean of each gdf. 
\subsubsection{Posterior Estimation} 
Having taken the necessary steps, we can now compute (\ref{eq:pgmf}) as an approximation for the posterior distribution in (\ref{eq:bayesfin}). Multiplying (\ref{eq:predpdf}) and (\ref{eq:mespdf}), the parameters of (\ref{eq:pgmf}) can be given by
\begin{align}
&\mug{\tetvec{i}}^k = \left[\sigg{\tetvec{i}}^k_-  + \sigg{\tetvec{i}}^k_+\right]^{-1} \left(\sigg{\tetvec{i}}^k_+\mug{\tetvec{i}}^k_- + \sigg{\tetvec{i}}^k_-\mug{\tetvec{i}}^k _+\right),
\label{eq:up11} \\
&\sigg{\tetvec{i}}^k = \left[\left(\sigg{\tetvec{i}}^k_-\right)^{-1}  + \left(\sigg{\tetvec{i}}^k_+\right)^{-1}\right]^{-1}.
\label{eq:up12}
\end{align}
The final estimation of the clock skew and offset can then be given by
\begin{align}
&\tg^k = \frac{1}{\mug{\tetvec{i}}^k[1]},& \ttet^k = \frac{\mug{\tetvec{i}}^k[2]}{\mug{\tetvec{i}}^k[1]}.&
\label{eq:finestpair}
\end{align}
Furthermore, each gdf can be updated across $\pos$ axis by
\begin{align}
&\mug{\pos}^k_f =  \left[\sigg{\pos}^k_{f-}  + \sigg{\pos}^k_{f+}\right]^{-1}\times  \left(\sigg{\pos}^k_{f+}\mug{\pos}^k_{f-} + \sigg{\pos}^k_{f-}\mug{\pos}^k_{f+}\right),
\label{eq:mu_est}\\
&\sigg{\pos}^k_f = \left[\left(\sigg{\pos}^k_{f-}\right)^{-1}  + \left(\sigg{\pos}^k_{f+}\right)^{-1}\right]^{-1}.
\label{eq:sig_est}
\end{align} 
Next, the weights can be updated as
\begin{equation}
w_f^k = \frac{w^k_{f-}w^k_{f+}}{\sum_{f=1}^F w^k_{f-}w^k_{f+}}.
\label{eq:weight_est}
\end{equation}
Given (\ref{eq:mu_est}), (\ref{eq:sig_est}), (\ref{eq:weight_est}), the final position estimation can be given by
\begin{equation}
\hat{\pos}^k = \sum_{f=1}^F w_f^k \mug{\pos}^k_f.
\label{eq:finwest}
\end{equation}
\subsubsection{Resampling and Tuning}
Resampling is one of the most crucial steps when performing PGM  filtering. Without the resampling step, the filter would suffer from sample depletion. That is, after a while all the gdfs but a few will have negligible weight. Consequently, the posterior will be approximated with only a few gdfs, leading to its underestimation. To overcome this shortcoming, in each iteration we replace the minor-weight gdfs with new ones whose means are sampled from the approximated posterior. The sample depletion can be monitored throughout the filtering process by calculating the number of effective gdfs as 
\begin{equation}
N_{\text{eff}} = \frac{1}{\sum_{f=1}^F (w_f^k)^2}.
\end{equation}
As can be seen, $N_{\text{eff}}$ attains its maximum when all the weights are equal to $\frac{1}{F}$ and falls to its minimum when all but a single weight is equal to zero. In this work, the resampling is carried out when the $N_{\text{eff}}< \frac{2}{3}F.$

All above-mentioned steps are summarized in algorithm~\ref{alg:brf}.

\subsection{Complexity of the Algorithm}
The computational complexity of different types of BRF and PF filters including L-BRF and PGM has been extensively discussed in \cite{gustafsson2010particle, karlsson2005complexity}. Parameters $l,$ $n,$ and $F$ denote the number of linear state variables, nonlinear state variables, and gdfs (or mixtures), respectively. For the sake of simplicity, we only consider the number of multiplications to evaluate the complexity. Table~\ref{tab:complexity} shows the complexity for each step of L-BRF and PGM. For the prediction step, it can be seen from (\ref{eq:predpdf}) and (\ref{eq:predpdfxy}) that two squared matrix multiplications and a matrix-vector multiplication are needed. We note that the computation cost of generating random variables is $O(1)$. The same holds for the likelihood computation given in (\ref{eq:musigzz}) and (\ref{eq:musigz}). In the PGM, however, the L-BRF is repeated $F$ times for each gdf across the nonlinear state variables. For the estimation step, the L-BRF needs 4 matrix inversions and 3 matrix-vector multiplications. The same number of multiplications is necessary for each gdf of the PGM. This is in addition to the multiplications between the weights and the particles essential to obtain the final estimation. Finally, we need to perform a cumulative sum to perform resampling, whose complexity is considered to be $O(F).$ It is apparent that PGM adds an overhead, however, it turns out that, according to \cite{gustafsson2010particle, karlsson2005complexity}, PGM is more efficient, especially when the uncertainty of the measurements increases.
\begin{table}[t!]
    \centering
    \caption{Complexity comparison of L-BRF and PGM filter.}
    \begin{tabular}{lcc}
        \hline
         & L-BRF & PGM \\ \hline
        Prediction & $2l^3 + l^2$ & $2l^3 + l^2$ \\
        Likelihood/Correction & $2l^3 + l^2$ & $2Fn^3 + Fn^2 + 2l^3 + l^2$ \\
        Estimation & $7l^2$ & $7Fn^2+nF+7l^2$ \\ \hline
        \textbf{Total} & $O(l^3)$ & $O(Fn^3+l^3)$ \\ \hline
    \end{tabular}
    \label{tab:complexity}
\end{table}

\begin{algorithm}[t!]
\begin{algorithmic}[1]
\STATE Initialize p($\tetvect{i}^0$) as in (\ref{eq:pgmf}).\label{init}
\FORALL {the APs in $\mathcal{I}_i$}  \label{forBRF}
\STATE Perform the time-stamp exchange mechanism described in Section \ref{sssec:schedule} and Figure \ref{fig:stamp}.
\STATE Estimate the CIR using QuaDRiGa channel model.
\STATE Estimate the AoA and the link condition $\zeta^{k}_{ij}$ using the CIR and (\ref{eq:aoa})
\FORALL {LoS links ($\zeta^{k}_{ij}$=0)}
\STATE Construct $\mathbf{B}_{ij}^k,$ $\mathbf{B}_{ij, f}^k,$ $\mathbf{R}_{ij}^k,$ $\mathbf{R}_{ij, f}^k,$ $\mathbf{r}_{ij}^k$ and $\mathbf{r}_{ij, f}^k$ by means of the time-stamps and the AoA.\label{mespdf} 
\STATE Update the parameters of the posterior distribution using (\ref{eq:up11}), (\ref{eq:up12}), (\ref{eq:mu_est}), and (\ref{eq:sig_est}).
\ENDFOR \label{convend}
\ENDFOR \label{endfor} 
\STATE Estimate the clock and position parameters using (\ref{eq:finestpair}) and (\ref{eq:finwest}).
\IF{$N_{\text{eff}} < \frac{2}{3}F$}
\STATE Perform resampling.
\ENDIF
\STATE Go to step \ref{forBRF}.
\end{algorithmic}
\caption{DePF joint sync\&loc.}
\label{alg:brf}
\end{algorithm}
\section{Simulation Results and Discussion}
In this section, we evaluate the performance of the techniques employed in this work. In particular, we first evaluate the performance of a DNN-based NLoS identifier. Next, we present the result of AoA estimation. Finally, the performance of the joint sync\&loc algorithm developed in this work is thoroughly analyzed.
\subsection{DNN-based NLoS Identification}
To perform NLoS identification, the DNN in Figure \ref{fig:dnn} needs to be trained first. The training data is obtained using the QuaDRiGa channel model, the details of which are given in \cite{jaeckel2014quadriga}. Specifically, the MU's movement profile can be implemented under the Urban Micro (UMi) cell scenario (denoted by ``3GPP$\_38.901\_$UMi" in the QuaDRiga documentation), which corresponds to the densely populated urban areas. We collect 5000 CIR realizations for each scenario, i.e., LoS and NLoS, 80$\%$ of which is used for the training purpose while the remaining 20$\%$ is treated as the test set. To prepare the CIRs to be fed into the DNN, we first input them into a 64-point FFT to obtain the Channel Frequency Responses (CFRs). Subsequently, we take the magnitude of the CFRs and normalize each to its maximum component so that all magnitudes are between $0$ and $1$. Such normalization is proved to result in faster learning and convergence \cite{shao2020normalization}. The normalized magnitudes of the CFRs are then fed into a DNN with $2$ hidden layers, each comprising $50$ neurons with a rectified linear unit activation function. The loss function in (\ref{eq:bce}) is then optimized using Adam optimizer to obtain the weights of each neuron. Furthermore, the probability that a CFR  corresponds to a LoS and NLoS link condition is indicated by the DNN's two output neurons with a softmax activation function. 

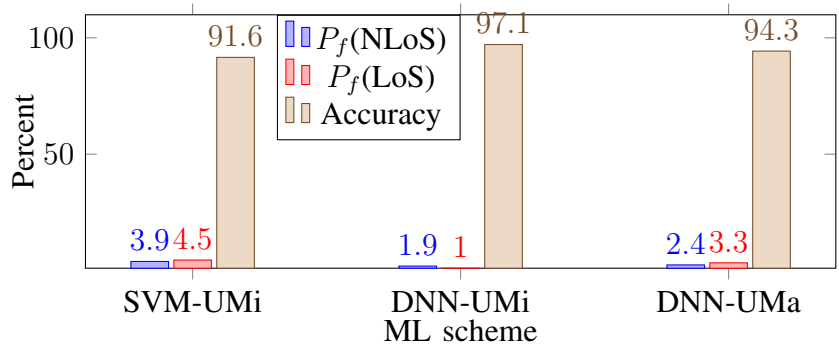
\begin{figure}
\begin{tikzpicture}
    \begin{axis}[
            ybar,
            ymax = 100,
            bar width=.5cm,
            enlarge x limits=0.2,
            enlarge y limits={value=.1,upper},
            xlabel = {ML scheme},
            xlabel style={yshift=-0.0cm},
            ylabel = {Percent},
            ylabel style={yshift=-.4cm},
            width=0.7\textwidth,
            height=.3\textwidth,
            legend style={at={(0.5, 1)}},
            symbolic x coords={SVM-UMi, DNN-UMi, DNN-UMa},
            xtick=data,
            nodes near coords,
        ]
        \addplot table[x=Scenario,y=$P_f$(NLoS)]{\mydata};
        \addplot table[x=Scenario,y=$P_f$(LoS)]{\mydata};
        \addplot 
        table[x=Scenario,y=Accuracy]{\mydata};
        \legend{$P_f$(NLoS), $P_f$(LoS), Accuracy}
    \end{axis}
\end{tikzpicture}
\centering
\caption{Comparison of two ML schemes when performing \mbox{NLoS-identification}. $P_f$(LoS)/$P_f$(NLoS) denotes the probability that the true condition of the links detected as LoS/NLoS is NLoS/LoS.}
\label{fig:nlos}
\end{figure}

Figure \ref{fig:nlos} depicts the accuracy of the NLoS-identifier based on SVM, a classical ML algorithm, and DNN, the method proposed in this work. As can be seen, the DNN-based method delivers higher accuracy, outperforming the classical method. Specifically, DNNs are more powerful when it comes to estimating the classifier function, and, therefore, they turn in superior performance. The performance remains high even if we employ the DNN in an environment other than that of the training data, i.e., Urban Macro (UMa) cells instead of UMi. If the environment is too dissimilar, the performance will drastically deteriorate. In our simulations, we observed a poor accuracy of $61\%$ for the rural-urban scenario, which is highly different from the UMi or UMa.

As mentioned before, the extremely high accuracy provided by the DNN is crucial as determining the link condition is among the most important decisions to be taken. In particular, false detection of NLoS links as LoS, $P_f$(LoS), not only can result in a poor estimation of the MU position and clock parameters, but also may lead to divergence of the filter. This occurs since the AoA estimation as well as the time-based distance measurement (which in the case of this work is carried out through time-stamp exchange) are highly inaccurate for NLoS links.
\pgfplotstableread[row sep=\\,col sep=&]{
Scenario & $P_f$(NLoS) & $P_f$(LoS) & Accuracy\\ 
SVM-UMi &  3.9 & 4.5 & 91.6 \\
DNN-UMi & 1.9 & 1.0 & 97.1 \\ 
DNN-UMa & 2.4 & 3.3 & 94.3 \\
}\mydata
\subsection{AoA estimation}
\begin{figure}[t!]
\includegraphics[width=.6\linewidth]{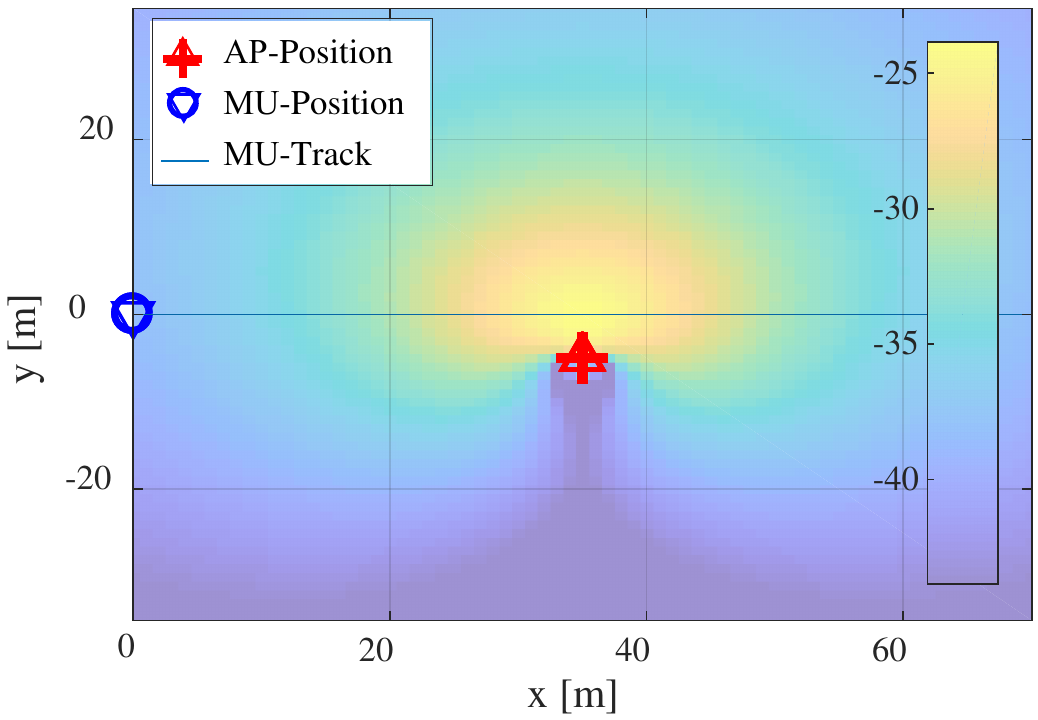}
\centering
\caption{Simulation setup for calculating the AoA.}
\label{fig:aoa-sim}
\end{figure}

To evaluate the performance of the MUSIC algorithm, we arrange a specific simulation setup (shown in Figure~\ref{fig:aoa-sim}) where an MU  moves with the velocity of $2$ m/s along the $x$ axis from the point $[x=0, y=0, z=1.5]$ until $[x=70, y=0, z=1.5].$ An AP with a $\nant\times \nant$ UPA and tilted 20$^{\circ}$ is located at $[x=35, y=-5, z=10],$ equally distant from the two edges of the trajectory. Figure~\ref{fig:aoa-sim} depicts such a setup where the MU's trajectory and AP's coverage area (for 23 dBm power allocated to each antenna element) are observable. Furthermore, the elements are assumed to be patch antennas with $90^{\circ}$ and $180^{\circ}$ beam opening in the elevation and azimuth plane, respectively. Such a setup covers all possible angles that an MU might have with respect to an AP, i.e., from $6^{\circ}$ to $171^{\circ}.$ Furthermore, it represents the basic movement of the MUs in an urban scenario, e.g., the movement profile of the users shown in Figure~\ref{fig:scenario} can be seen as the combination of that depicted in Figure~\ref{fig:aoa-sim}. Lastly, at each time step, the AoA is estimated using the MUSIC algorithm fed with the corresponding CIR generated by QuaDRiGa. The algorithm estimates the azimuth and elevation AoA using the binary exhaustive search up to the 0.5 degree accuracy level, where the number of search bins are $40$ and $20$, respectively.

Figure~\ref{fig:aoa} depicts the Root Mean Square Error (RMSE) of the AoA estimation for several UPA sizes. As can be observed, the RMSE of azimuth AoA estimation remains under $1.5^{\circ}$ for almost all the investigated UPAs, which paves the way for precise localization of the MUs. Nevertheless, in our simulations, we observed that for smaller UPAs the RMSE increases drastically due to the large errors at the edges of the trajectory. Although such cases rarely occur, they can potentially lead to filter divergence. Moreover, the same behavior is observed for the elevation AoA estimation. Generally, as can be seen in the figure, the RMSE is slightly higher for the elevation AoA since the MU is always in the $[10^{\circ}-50^{\circ}]$ angle sight of the AP. We know that UPA's estimation performance deteriorates as we move towards the edges. In practice, due to the density of the APs, the MUs are expected to be in the azimuth angle range of $[20^{\circ} - 150^{\circ}],$ and in the elevation angle range of $[20^{\circ}-50^{\circ}],$ i.e., AP density of fewer than 60 meters.
\pgfplotstableread[row sep=\\,col sep=&]{
Scenario & Azimuth & Elevation \\ 
$3\times 3$ & 1.32 & 2.74 \\
$4\times 4$ & 1.19 & 1.87\\
$5\times 5$ & 1.11 &1.80\\
$6\times 6$ & 1.07  & 1.44\\
}\mydata
\begin{figure}[t!]
\begin{tikzpicture}
    \begin{axis}[
            ybar,
            ymin = 0,
            ymax = 2.8,
            enlarge x limits=0.15,
            enlarge y limits={value=.2,upper},
            xlabel = {UPA size.},
            xlabel style={yshift=-0.2cm},
            ylabel = {RMSE of AoA estimation ($^\circ$).},
            ylabel style={yshift=-.5cm},
            bar width=.5cm,
            width=0.6\textwidth,
            height=.3\textwidth,
            legend style={at={(0.5,1)},
                anchor=north,legend columns=-1},
            symbolic x coords={$3\times 3$,$4\times 4$, $5\times 5$, $6\times 6$},
            xtick=data,
            nodes near coords,
        ]
        \addplot table[x=Scenario,y=Azimuth]{\mydata};
        \addplot table[x=Scenario,y=Elevation]{\mydata};
        \legend{Azimuth, Elevation}
        
    \end{axis}
\end{tikzpicture}
\centering
\caption{AoA estimation accuracy.}
\label{fig:aoa}
\end{figure}
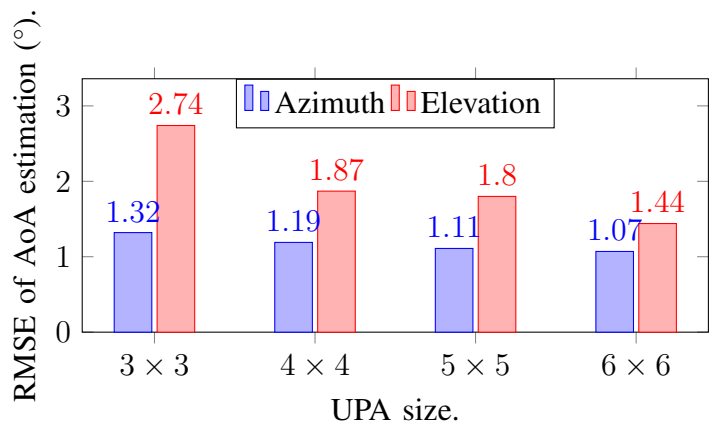
\begin{table}[t!]
\centering
\caption{Simulation parameters}
\begin{tabular}{lc}
\hline
General parameters & Values \\ \hline
\# of independent simulations &	1000 \\ 
Initial random delays ($\tilde{\theta}_i$) &	$\mathcal{U}(-10^3, 10^3)$ ns \\  
Initial random skew ($\gamma_i$) &	$\mathcal{U}(1-10^{-4}, 1+10^{-4})$ or $[0-100]$ ppm \\ 
Max. MU velocity & 14 (m/s) \\
AP density & 50 m \\ 
Distance traversed by the MU & 600 m\\ \hline
QuaDRiGa parameters & \\ \hline
Scenario & 3GPP$\_$38.901$\_$UMi \\
Center Frequency / FFT size ($N_s$) & 3.8 GHz / 64 \\
\# of MU/AP antenna ($N_{\text{ant}}$) & 1 / $3\times 3$ \\ \hline
Filter Parameters &  \\ \hline
Period of joint sync\&loc ($T$) & 100 ms \\
Process noise covariance matrix ($\mathbf{Q}_n$) & $\diag(10^{-5}, 10, 1.5, 1.5)$ \\ 
\# of Gaussian mixtures (gdfs) & 500 \\ \hline
 DNN parameters & \\ \hline 
$l_H, $ $n_H$ & 2, 50 \\
Optimizer & Adam (lr$=0.001$, beta$\_$1=$0.9$, beta$\_$2=$0.999$) \\
\# of epochs & 10\\
Batch size & 16 \\
Activation function of hidden layers & ReLU \\
Activation function of output layer & Softmax
\\ \hline
\end{tabular}
\label{tab:sim}
\end{table}
\subsection{Joint sync\&loc}\label{ssec:scenario}
We perform analysis for the scenario shown in Figure~\ref{fig:scenario}, which is regarded in \cite{werner2015joint, koivisto2017joint} as challenging. A car commences its journey by accelerating to reach the velocity of $14$ m/s ($= 50$ km/h). It continues moving with constant velocity and decelerates upon approaching the intersection until it completely stops (e.g., due to the red light). The same repeats between the two intersections. At the second intersection, it begins moving, then takes a turn, and continues to accelerate to $14$ m/s limit until it exits the map. All the turns, as well as the acceleration coefficients, are chosen randomly. During its journey, at each joint sync\&loc round $k,$ the MU exchanges time-stamps with a fixed number of APs ($N_{\text{AP}}$) in $\mathcal{I}_i,$ the link to each of which is LoS/NLoS with the probability of $0.8$/$0.2.$ The APs are grouped into $\mathcal{I}_i$ based on the distance criteria, that is, $\mathcal{I}_i$ includes the $N_{\text{AP}}$ closest APs to the $i$-th MU. A further assumption is that, at each joint sync\&loc period $T,$ $\nant\times\nant$ CIRs are available at each AP connected to the MU. In our simulations, the CIRs are obtained using the QuaDRiGa channel model. More explicitly, at each round $k$, knowing the true MU-AP distance and the link condition, i.e., LoS or NLoS, the CIRs are generated using the ``3GPP$\_$38.901$\_$UMi" scenario of the QuaDRiGa channel model. Moreover, the RMSEs obtained by \cite{werner2015joint, koivisto2017joint} serve as the baseline to our approach. The second scheme with which we compare our proposed algorithm is the L-BRF filtering proposed in \cite{goodarzi2020bayesian, goodarzi2021synchronization}. The aforementioned approaches are the most relevant as they draw on the same inputs as our proposed method does.
 
We initialize all the clock offsets from the  $\mathcal{U}(-10^3, 10^3)$ ns. The initial skews of all the clocks are drawn from the uniform distribution $\mathcal{U}(1-10^{-4}, 1+10^{-4}),$ which corresponds to skew values between 0 and 100 \textit{part-per-million} (ppm). The covariance of the clock process noise $\qn{\tetvec{i}}$ is set to $\diag(10^{-5}, 100)$ to account for the residual errors from the previous iterations as well as the external noises on the clock skew and offset. The covariance of position process noise $\qn{\pos}$ amounts to $\diag((14T)^2, (14T)^2)$ to account for every possible movement of the MU.  
All the additional simulation parameters can be found in Table \ref{tab:sim}. 

\begin{figure}[t!]
\begin{tikzpicture}
    \begin{axis}[
            ybar,
            ymax = 10,
            ymin = 0,
            enlarge x limits=0.15,
            enlarge y limits={value=.1,upper},
            xlabel = {\# of APs},
            xlabel style={yshift=-0.2cm},
            ylabel = {Clock offset RMSE (ns)},
            ylabel style={yshift=-.5cm},
            bar width=.5cm,
            width=0.6\textwidth,
            height=.3\textwidth,
            legend style={at={(0.5,1)},
                anchor=north,legend columns=-1},
            symbolic x coords={1-AP LoS, 2-AP LoS, 3-AP LoS, 3-AP },
            xtick=data,
            nodes near coords,
        ]
        \addplot table[x=Scenario,y=EKF]{\mydata};
        \addplot table[x=Scenario,y=L-BRF]{\mydata};
        \addplot table[x=Scenario,y=DePF]{\mydata};
        \legend{EKF, L-BRF, DePF}
    \end{axis}
   \node (a) at (.3, 3.5){\color{blue}{$33$}};
\end{tikzpicture}
\centering
\caption{Performance comparison of three joint synchronization and localization algorithms in terms of clock offset estimation.}
\label{fig:compclk}
\end{figure}
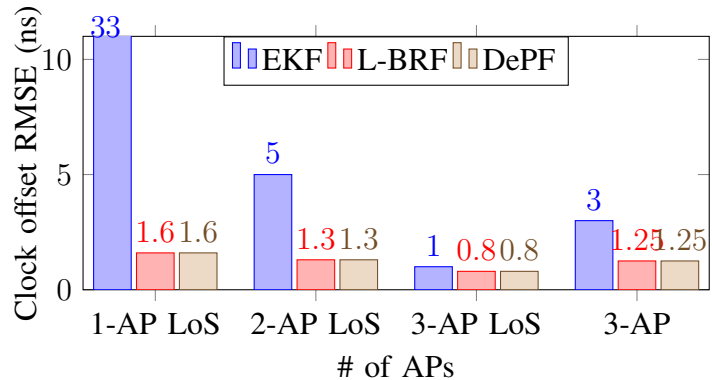

Figure \ref{fig:compclk} shows the RMSE of clock offset estimation for three joint snyc\&loc algorithms. The DePF algorithm is compared with two linear Bayesian methods, i.e., EKF and L-BRF, in multiple scenarios. In particular, we compute the RMSEs in three scenarios, with the number of LoS APs ranging from 1 to 3. In an additional scenario, we consider the MU being connected to three APs, where each MU-AP link condition is set to LoS with the probability of 0.8. As can be seen, for all the LoS scenarios, the L-BRF and DePF deliver an identical performance, which is expected as they rely on the same approach to estimate the clock parameters. On the other hand, the performance of the EKF falls behind as it does not explicitly draw on the synchronization signals to estimate the clock offset. Moreover, the synchronization algorithm scheme utilized to synchronize the APs, i.e., hybrid BP-BRF network synchronization, leads to a more precise inter-AP synchronization and, consequently, it lowers the MU clock offset estimation error. In the last case, the L-BRF and DePF that draw on DNN-based NLoS identification outperform the EKF-based method where the NLoS links are identified by means of Rice factor of the incoming signal strength.
\pgfplotstableread[row sep=\\,col sep=&]{
Scenario & EKF & L-BRF & DePF \\ 
1-AP LoS & 33 & 1.6 & 1.6 \\
2-AP LoS & 5 & 1.3 & 1.3 \\
3-AP LoS & 1 & 0.8 & 0.8 \\
3-AP   & 3 & 1.25 & 1.25 \\ 
}\mydata

\begin{figure}[t!]
\begin{tikzpicture}
    \begin{axis}[
            ybar,
            enlarge x limits=0.15,
            ymin=0,
            ymax=4.3,
            xlabel = {\# of APs},
            xlabel style={yshift=-0.2cm},
            ylabel = {Position RMSE (m)},
            ylabel style={yshift=-.5cm},
            bar width=.5cm,
            width=0.6\textwidth,
            height=.3\textwidth,
            legend style={at={(0.5,1)},
                anchor=north,legend columns=-1},
            symbolic x coords={1-AP LoS, 2-AP LoS, 3-AP LoS, 3-AP },
            xtick=data,
            nodes near coords,
        ]
        \addplot table[x=Scenario,y=EKF]{\mydata};
        \addplot table[x=Scenario,y=L-BRF]{\mydata};
        \addplot table[x=Scenario,y=DePF]{\mydata};
        \legend{EKF, L-BRF, DePF}
    \end{axis}
\end{tikzpicture}
\centering
\caption{Performance comparison of three joint synchronization and localization algorithms in terms of position estimation.}
\label{fig:comppos}
\end{figure}
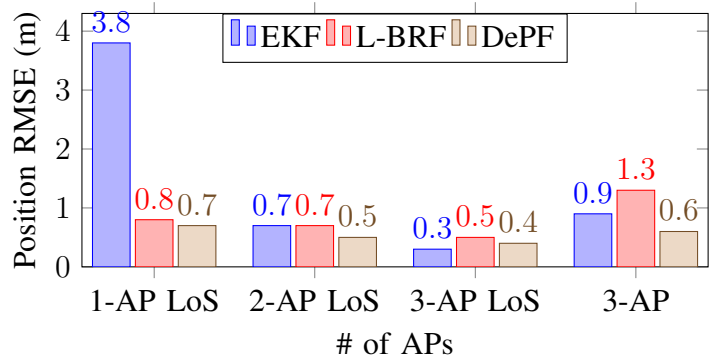

Figure \ref{fig:comppos} depicts the RMSE of position estimation for three joint sync\&loc algorithms. The DePF algorithm is compared with two linear Bayesian methods, i.e., EKF and L-BRF, in the same scenarios as in Figure~\ref{fig:compclk}. As can be seen, for almost all the scenarios, the DePF algorithm delivers superior performance. In particular, since the DePF employs a higher number of gdfs, rather than only one, to approximate the posterior distribution it can estimate the position more accurately. Furthermore, DePF stands out when dealing with NLoS links. This is straightforward to notice as the RMSE of position estimation is lower for DePF in the 3-AP scenario where the L-BRF employs the same NLoS identifier as DePF. Additionally, unlike EKF and L-BRF, DePF does not need any initialization, which is of crucial importance in practice as initialization would require the APs to request position estimation from the MUs, which may not be always possible. Overall, considering 2-AP LoS, 3-AP LoS, and 3-AP scenarios, EKF and L-BRF perform close to DePF when both a reliable initialization and MU-AP links with known LoS conditions are available. Nevertheless, such assumptions are questionable in practice, rendering the EKF-based and L-BRF algorithms futile in real-world scenarios.
\pgfplotstableread[row sep=\\,col sep=&]{
Scenario & EKF & L-BRF & DePF  \\ 
1-AP LoS & 3.8 & 0.8 & 0.7  \\
2-AP LoS & 0.7 & 0.7 & 0.5 \\
3-AP LoS & 0.3 & 0.5 & 0.4 \\
3-AP  & 0.9 & 1.3 & .6 \\ 
}\mydata

\begin{figure}[t!]
\includegraphics[width=.6\linewidth]{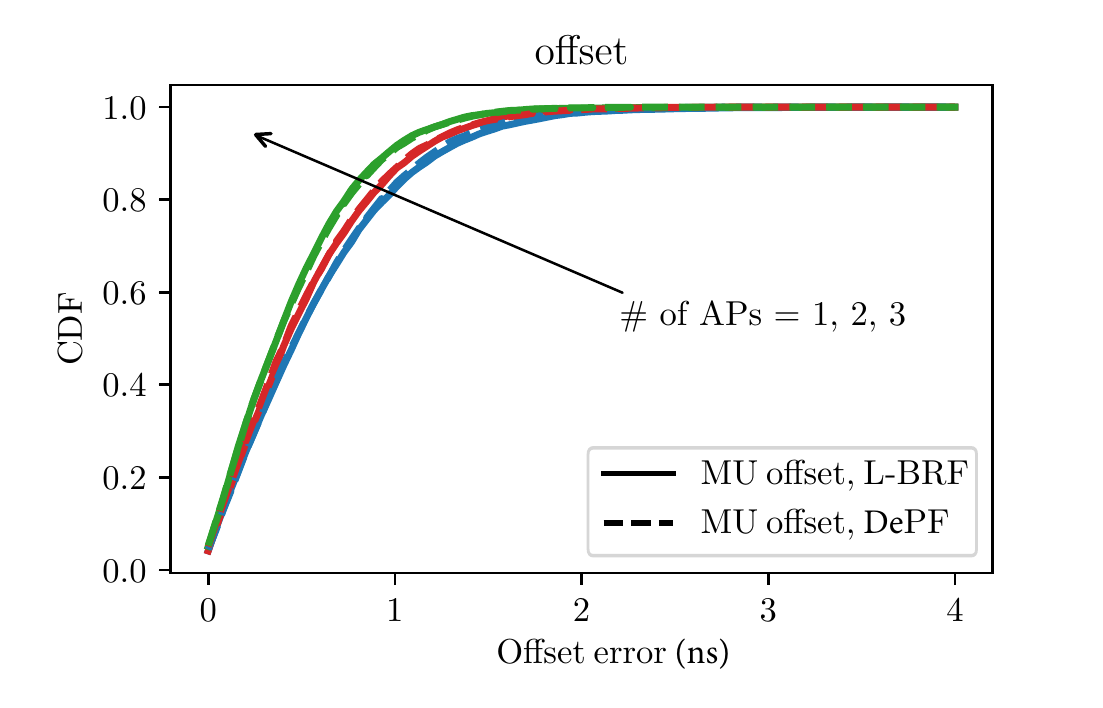}
\centering
\caption{Performance comparison of L-BRF and DePF when estimating the MUs' clock offset.}
\label{fig:perform1}
\end{figure}

Hereafter, all the simulations have been carried out assuming that there is always at least one LoS MU-AP link. Figure~\ref{fig:perform1} presents the CDF of the clock offset estimation error when the MU is connected to multiple APs. It can be seen that the estimation accuracy always remains below 2 ns and increases as both L-BRF and DePF utilize more measurements to estimate the clock offset and skew. In fact, since the APs are synchronized with high precision, collecting time-stamps from each additional AP does provide additional information about the statistics of MU's clock parameters and, therefore, increases the accuracy of the estimation. Such precision is necessary if the location of the MU is to be accurately estimated. We note that each single ns inaccuracy maps to 0.3 m distance measurement error and, consequently, worsens the location estimation. Furthermore, the performance of both schemes is identical as they draw on the same approach, i.e., modeling the clock parameter with a single gdf, to estimate the clock parameters.

\begin{figure}[t!]
\includegraphics[width=.6\linewidth]{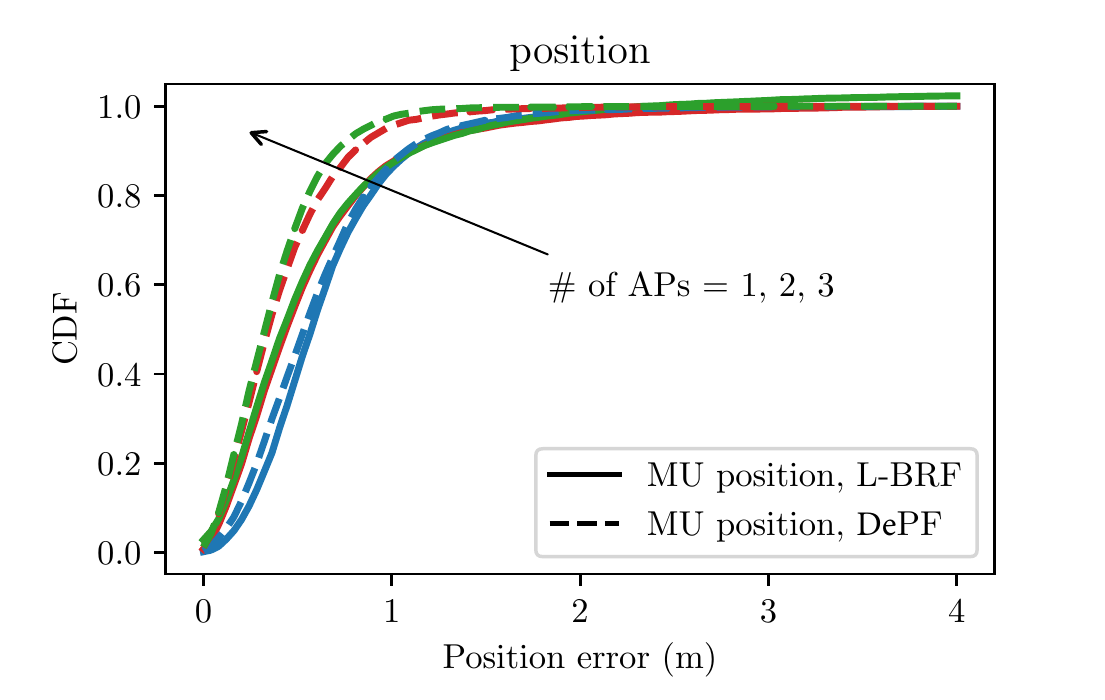}
\centering
\caption{Performance comparison of L-BRF and DePF when estimating the MUs' position.}
\label{fig:perform2}
\end{figure}

Figure~\ref{fig:perform2} presents the CDF of the position estimation error when the MU is connected to multiple APs. As can be seen, the position estimation error is less than 1 meter in $90\%$ of the cases for the DePF algorithm. We observe that DePF significantly outperforms the L-BRF, especially for 2- and 3-AP scenarios. In particular, unlike the L-BRF that approximates the posterior with a single Gaussian distribution, in DePF, the approximation is based on multiple gdfs. Consequently, the approximated posterior is closer to the true one, leading to a more precise position estimation. Another subtle observation is that, although the position estimation error decays with the growth in the number of APs, increasing the number of APs from 2 to 3 only slightly improves the performance. In fact, the third AP is normally far away from the MU, leading to a poorer (AoA and time-stamp) measurement accuracy compared to that of the first two APs. Hence, it does not provide substantial further information about the posterior distribution of the MU's location.   

\begin{figure}[t!]
\includegraphics[width=0.6\linewidth]{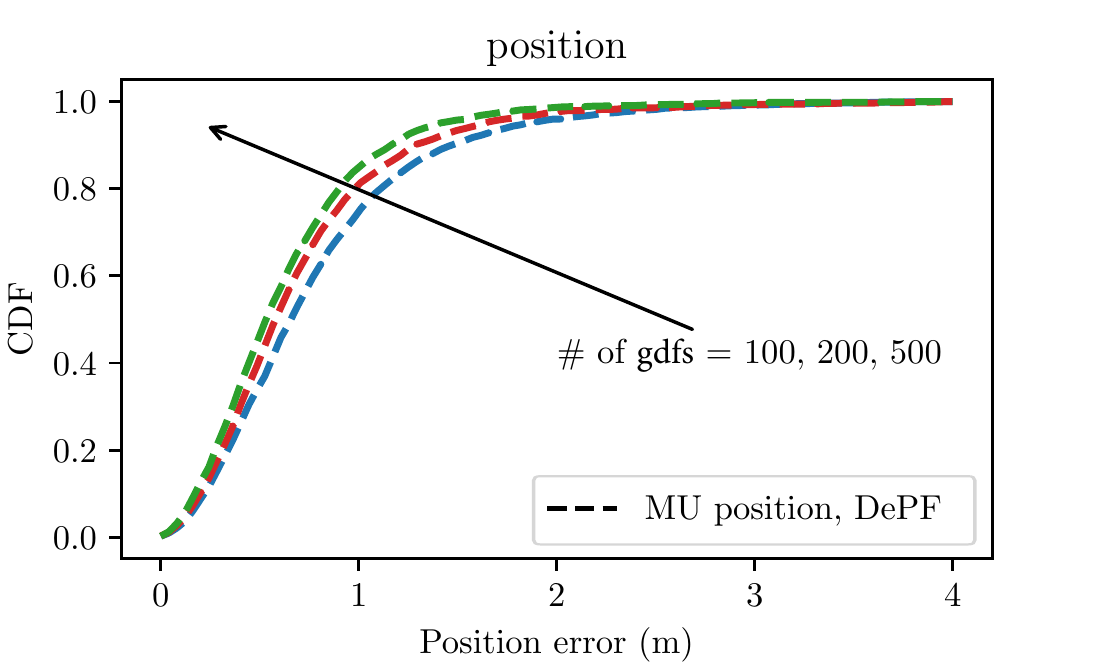}
\centering
\caption{Performance of joint sync\&loc algorithm for different number of gdfs.}
\label{fig:perform3}
\end{figure}

Figure \ref{fig:perform3} indicates the CDF of the position estimation for multiple numbers of gdfs. It can be noticed that the position estimation ameliorates with the increase of the number of gdfs. This is expected as in PGM filters the posterior distribution is approximated by multiple gdfs. Consequently, the more gdfs we employ, the more accuracy we achieve, albeit with higher computation time. Nevertheless, the error reduction is decreasing when increasing the number of gdfs, suggesting that a proper balance needs to be struck between the number of gdfs and the localization accuracy. In the scenarios presented in this work, one can achieve satisfactory performance even with 500 gdfs.

\begin{figure}[t!]
\includegraphics[width=0.6\linewidth]{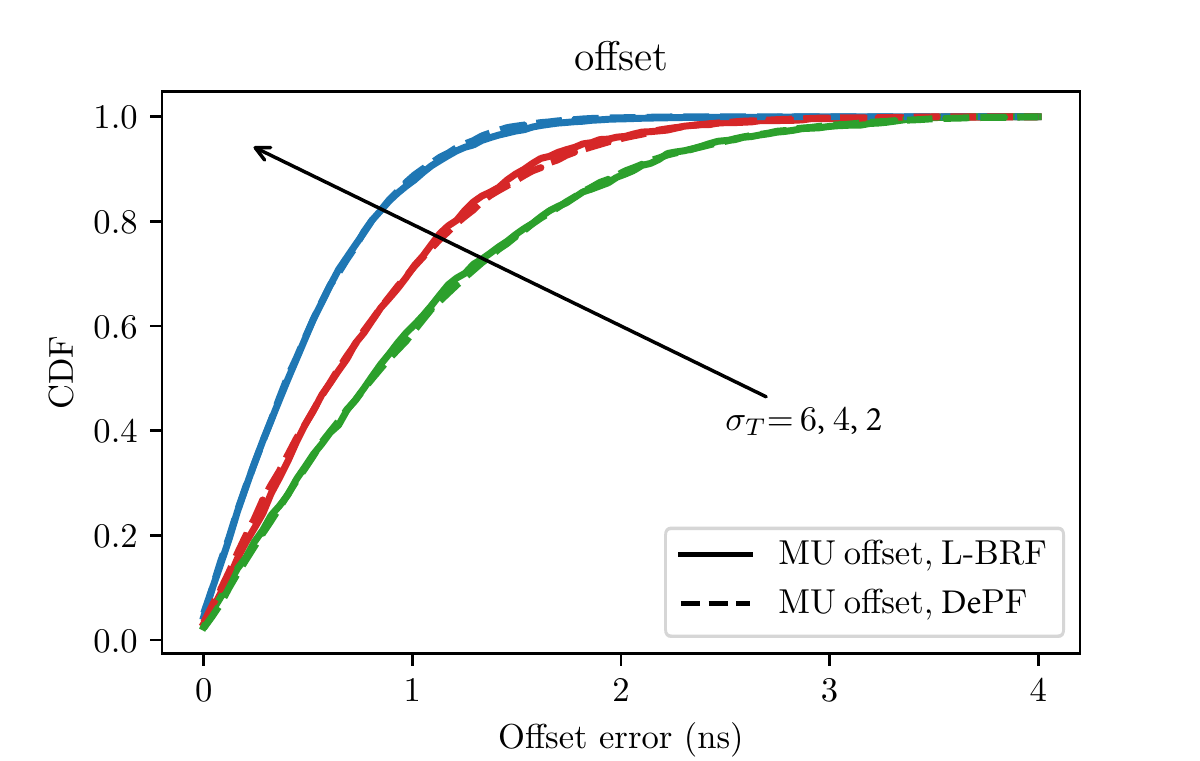}
\centering
\caption{Clock offset estimation performance of joint sync\&loc algorithm with different number of APs involved.}
\label{fig:perform4}
\end{figure}

Figure \ref{fig:perform4} shows the CDF of the clock offset estimation error carried out by a single AP for different time-stamp uncertainties, i.e., $\sigma_T = 2, 4, 6.$ As can be seen, the clock offset estimation accuracy drops as the $\sigma_T$ grows. It remains, however, less than 3 ns in $90\%$ of the cases. Such degradation can cause an additional error in position estimation as, given (\ref{eq:c3}), both parameters are intertwined. Specifically, the offset estimation error can introduce distance measurement error, resulting in imprecision when estimating the position.  Nevertheless, the uncertainty of the time-stamping of the state-of-the-art devices is expected to be below 5 ns. Moreover, the destructive impact of the uncertainty can be also mitigated by employing more synchronized APs as discussed previously and shown in Figure~\ref{fig:perform1}.

\begin{figure}[t!]
\includegraphics[width=0.6\linewidth]{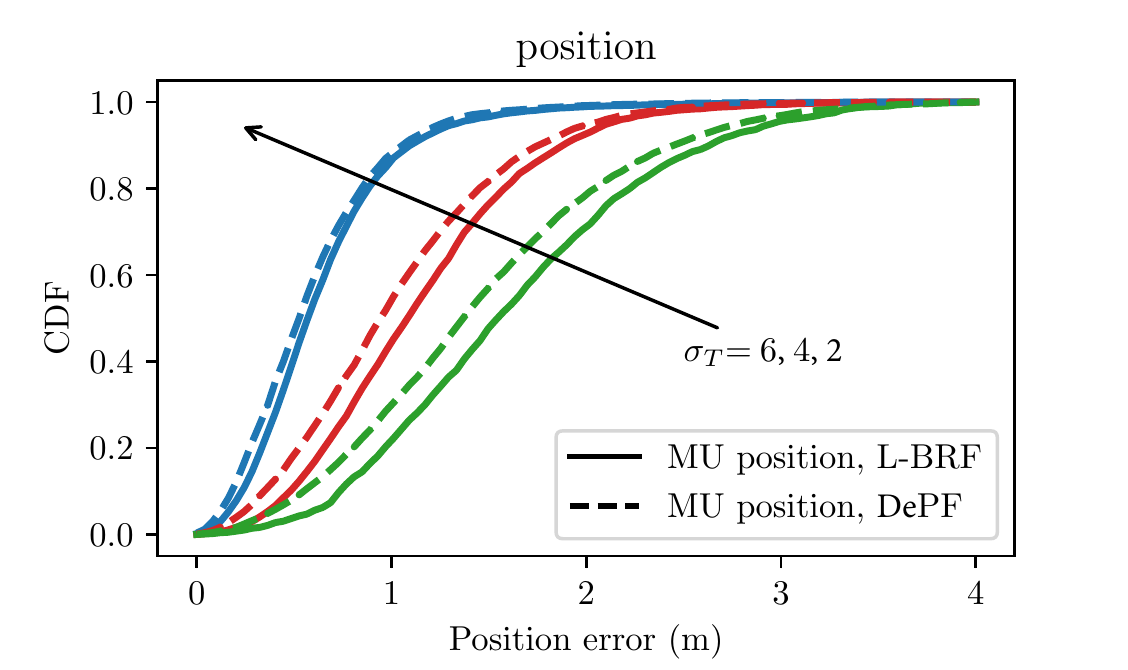}
\centering
\caption{Position estimation performance of joint sync\&loc algorithm with different time-stamp accuracy.}
\label{fig:perform5}
\end{figure}

Figure \ref{fig:perform5} shows the CDF of position estimation conducted by a single AP for different time-stamp accuracies. It can be noticed that the position estimation accuracy deteriorates with the growth in the time-stamp uncertainty. Specifically, the growth in uncertainty results in more erroneous distance measurements and offset estimations, which, consequently, worsens the position estimation accuracy. Nevertheless, it can be readily seen that DePF is more successful in mitigating the destructive effect of the time-stamp uncertainty. Moreover, for both DePF and L-BRF, employing more APs can alleviate the negative impact of large time-stamp uncertainty. In both Figures~\ref{fig:perform4} and \ref{fig:perform5}, it can be noticed that $\sigma_T$ plays a decisive role in the outcome of the estimation algorithm, which also reveals the importance of hardware components in the design of a robust and precise joint sync\&loc algorithm. In practice, such uncertainty in commercial off-the-shelf devices is expected to be below 5 ns.

In summary, one can see that DNNs can play a decisive role by facilitating accurate decision-making in simple, albeit crucial, tasks such as NLoS identification. Furthermore, it can be noticed that in the case when we have multiple LoS links available, the performance of the EKF-based and L-BRF approaches both in terms of clock offset and position is close to that of DePF. Nevertheless, in the absence of LoS condition, DePF demonstrates more competence in estimating the clock and position parameters by employing only a few hundred gdfs. Another point worth mentioning is that time-stamp exchange is of high potential to be employed for performing joint sync\&loc. In particular, the current communication devices are capable of performing FTM up to 5 ns accuracy, fertilizing the ground for precise offset and distance measurements, which are the basis for precise joint synchronization and localization.
\section{Conclusion and Future works}
We presented a DNN-assisted Particle-based filtering (DePF) algorithm for joint synchronization and localization (sync\&loc) of Mobile Users (MUs) in communication networks. In particular, we leveraged an asymmetric time-stamp exchange mechanism, traditionally utilized for time synchronization, to estimate the clock offset and skew while simultaneously obtaining information about the distance between the access points and the MUs. Further on, we combined the aforementioned measurements with the angle of arrival estimation and the link condition, i.e., line-of-sight or non-line-of-sight, returned by a pretrained DNN to localize the MUs. Simulation results indicate that while the performance of the proposed algorithm is promising, especially under challenging real-world conditions, the position and clock offset estimation errors are dependent on the accuracy of hardware time-stamping. We mitigated the negative impact of this dependency by deploying more access points for performing joint sync\&loc. 

In this work, we drew on simulations to prove the efficiency of our proposed algorithm. However, to cross-validate the obtained results, the algorithm needs to be implemented in practice. Therefore, in future works, we will employ the hardware at our disposal to evaluate the performance of our proposed joint sync\&loc algorithm in practice.
\section*{Abbreviations}
\abb{AoA}{Angle of Arrival}, \abb{AP}{Access Point}, \abb{BN}{Bayesian Network}, \abb{BP}{Belief Propagation}, \abb{BS}{Base Station}, \abb{CDF}{Cumulative Distribution Function}, \abb{CIR}{Channel Impulse Response}, \abb{CFR}{Channel Frequency Response}, \abb{DePF}{DNN-assisted Particle-based Bayesian Filtering}, \abb{DNN}{Deep Neural Network}, \abb{EKF}{Extended Kalman Filter}, \abb{FFT}{Fast Fourier Transform}, \abb{FTM}{fine time measurement}, \abb{gdf}{Gaussian density function}, \abb{L-BRF}{Linearized Bayesian Recursive Filtering}, \abb{LoS}{Line-of-Sight}, \abb{ML}{Machine Learning}, \abb{MU}{Mobile User}, \abb{NLoS}{Non-Line-of-Sight}, \abb{PGM}{Particle Gaussian Mixture}, \abb{RMSE}{Root Mean Square Error}, \abb{sync\&loc}{Synchronization and Localization}, \abb{SVM}{Support Vector Machine},  \abb{UPA}{Uniform Linear Array}.
\balance
\bibliography{synch_loc_j}
\bibliographystyle{IEEEtran}
\end{document}